\documentclass[linenumbers,twocolumn]{aastex701}
\usepackage{placeins}

\usepackage{tikz}
\usetikzlibrary{arrows.meta, positioning, shapes.geometric}
\usepackage{algorithm}
\usepackage{algpseudocode}
\usepackage{booktabs}
\usepackage{multirow}
\usepackage{amsmath}

\begin{document}


\title{Syntriod: A Robust Initial Parameter Estimator for Radial Velocity Curve Solutions Beyond Conventional Sampling Limits}

\author[0000-0002-3413-650X]{*Emre Barbaros}
\affiliation{Department of Astronomy and Space Sciences, Faculty of Science, Erciyes University, 38039, Kayseri, Türkiye}
\affiliation{Department of Astronomy and Space Sciences, Graduate School of Natural and Applied Sciences, Erciyes University, 38039, Kayseri, Türkiye}
\affiliation{Astronomy and Space Sciences Observatory and Research Center, Erciyes University, Kayseri, Türkiye}
\email{ebabaros054@gmail.com}

\author[0000-0001-9385-5005]{Hasan Ak}
\affiliation{Department of Astronomy and Space Sciences, Faculty of Science, Erciyes University, 38039, Kayseri, Türkiye}
\affiliation{Department of Astronomy and Space Sciences, Graduate School of Natural and Applied Sciences, Erciyes University, 38039, Kayseri, Türkiye}
\affiliation{Astronomy and Space Sciences Observatory and Research Center, Erciyes University, Kayseri, Türkiye}
\email{hasanak99@gmail.com}

\author[0000-0003-3016-5490]{N. Filiz Ak}
\affiliation{Department of Astronomy and Space Sciences, Faculty of Science, Erciyes University, 38039, Kayseri, Türkiye}
\affiliation{Department of Astronomy and Space Sciences, Graduate School of Natural and Applied Sciences, Erciyes University, 38039, Kayseri, Türkiye}
\affiliation{Astronomy and Space Sciences Observatory and Research Center, Erciyes University, Kayseri, Türkiye}
\email{nfilizak@gmail.com}

\email[show]{ebarbaros054@gmail.com}

\begin{abstract}

We present \texttt{Syntriod}, an orbital-phase domain RV template-based algorithm designed to provide robust initial orbital parameter estimates for spectroscopic binaries across a wide range of observational sampling conditions. Rather than performing full orbital inference, \texttt{Syntriod} aims to constrain the parameter space with physically consistent solutions that can guide subsequent optimization procedures.
We evaluate its performance using 10,000 synthetic Keplerian orbits spanning diverse orbital configurations and sampling regimes. For well-sampled datasets ($N_{\mathrm{obs}} \geq 8$), \texttt{Syntriod} recovers orbital periods with relative accuracies of order $\sim 10^{-3}$. At the theoretical sampling limit ($N_{\mathrm{obs}} = 6$), the method maintains a success rate of {$\sim 94\%$}, while classical period-search techniques such as Lomb–Scargle become increasingly affected by aliasing. Even below this limit, 
\texttt{Syntriod} continues to recover the correct orbital solution in {$\sim 83\%$} of cases with $N_{\mathrm{obs}} = 5$, showing a gradual degradation in precision rather than catastrophic failure.
We further evaluate the method on 12 real spectroscopic binary systems (HD 160934, Phi~Cyg, Capella~A, Kepler~16, KIC 3858884, KIC~6867766, KIC~2445134, KIC~3003991, DU~Boo, HL~Dra, FP~Boo, and AK~Her) spanning a broad range of orbital periods and eccentricities. \texttt{Syntriod} consistently reproduces literature solutions even when datasets are randomly subsampled to sparse regimes. In cases where a full Keplerian solution becomes underconstrained ($N_{\mathrm{obs}} \leq 4$), the algorithm transitions to linear dynamical relations and recovers parameters such as the mass ratio ($q$) and systemic velocity ($\gamma$) with success rates exceeding 99\%.
These results demonstrate that \texttt{Syntriod} provides reliable and computationally efficient initial parameter estimates across both well-sampled and sparse observational regimes, making it a practical pre-solver for modern orbit-fitting pipelines and large spectroscopic surveys.

\end{abstract}



\section{Introduction}
\label{sec:Intro}

Radial velocity (RV) measurements remain one of the primary tools for determining the orbital properties of spectroscopic binary systems and exoplanet hosts. The rapid growth of large-scale surveys such as \textit{GAIA } \citep{gaia}, {SDSS \citep{SDSS}}, {LAMOST \citep{Lamost}} and \textit{TESS} \citep{Tess}, along with upcoming missions like \textit{PLATO} \citep{plato}, has significantly increased the number of targets requiring spectroscopic characterization. However, for many of these systems, the available RV data are limited, often consisting of a small number of measurements obtained at irregular time intervals. Under such conditions, deriving reliable orbital parameters becomes a challenging inference problem.

Modern orbital fitting frameworks such as \texttt{rvfit} \citep{rvfit}, \texttt{RadVel} \citep{radvel}, \texttt{juliet} \citep{julier}, and \texttt{BinaryStarSolver} \citep{BSSolver} typically rely on Bayesian inference techniques, including Markov Chain Monte Carlo (MCMC) sampling \citep{markovchain, montecarlo} or Nested Sampling \citep{nested}. These approaches are powerful tools for exploring complex parameter spaces and estimating uncertainties. However, their efficiency strongly depends on the availability of reasonable initial parameter estimates. In practice, most orbital solvers assume that the orbital period is already known to within a narrow range. In particular, although \texttt{rvfit} employs Adaptive Simulated Annealing (ASA) \citep{ASA}—a powerful optimization technique—even this robust routine recommends keeping the orbital period fixed or restricting the search to a very narrow interval, rather than scanning a broad parameter space. If this initial estimate is inaccurate, the algorithm may converge toward incorrect solutions or require prohibitively long computation times.


Consequently, determining reliable preliminary orbital parameters remains a critical step in the analysis pipeline. In many cases, this initial exploration is performed using period-search techniques such as the Lomb-Scargle (LS) periodogram \citep{lomb}. While LS is widely used due to its computational efficiency, it assumes a sinusoidal signal model and therefore performs best for nearly circular orbits. In eccentric systems or datasets with irregular temporal sampling, the signal power is often distributed across multiple harmonic frequencies, increasing the risk of aliasing and incorrect period identification. Alternative methods operating in the phase-folded time domain, such as String Length \citep[SL;][]{SL} and Phase Dispersion Minimization \citep[PDM;][]{PDM}, provide greater flexibility but tend to become unstable in low-sampling regimes.

The difficulty becomes more fundamental when the number of observations approaches the dimensionality of the orbital parameter space. For a single-lined spectroscopic binary (SB1) with a general Keplerian orbit, the solution is described by six parameters, 
${\theta = (P, e, \omega, T_0, K, \gamma)}$  \citep{Hilditch}. This limitation implies that at least 6 independent RV measurements are required to formally constrain the system. The limit decreases to five for circular orbits and increases to seven for double-lined systems (SB2). In practice, however, solutions obtained near this theoretical threshold remain highly sensitive to noise and phase coverage. In this work, we adopt $N_{\mathrm{obs}} = 6$ as the effective sampling limit and define datasets with $N_{\mathrm{obs}} \geq 8$ as well-sampled. Below this limit, the parameter space becomes increasingly underconstrained, and classical period-search methods tend to fail due to aliasing, while full Bayesian approaches become computationally inefficient due to the large parameter space to be explored.

In this study, we present \texttt{Syntriod}, 
an orbital-phase domain RV template-based algorithm designed to provide reliable initial orbital parameter estimates across both well-sampled and sparsely sampled datasets. Instead of performing full orbital inference, the method constrains the parameter space by directly comparing observations to a library of precomputed Keplerian radial-velocity templates. This approach enables the simultaneous estimation of orbital parameters while maintaining computational efficiency.
The details of the algorithm are in Section~\ref{sec:method}.
We evaluate the performance using both synthetic datasets spanning a wide range of orbital configurations and 12 real spectroscopic binary systems drawn from the literature in Section \ref{sec:test_results}. Our results highlight regimes where classical period-search techniques struggle, including highly eccentric systems and datasets with fewer than 6 observations.
In addition to providing robust period estimates, the algorithm exploits the coupled dynamics of double-lined spectroscopic binaries to extract physically meaningful parameters, such as the mass ratio ($q$) and systemic velocity ($\gamma$), even with only 2 RV observations.  
By combining deterministic template matching with auxiliary validation strategies for sparse data regimes, \texttt{Syntriod} provides reliable initial orbital parameters that can significantly reduce the computational cost of subsequent Bayesian orbit-fitting procedures.

\section{Methodology of \texttt{Syntriod}: An Orbital Phase-Domain RV Template-Matching Approach} \label{sec:method}

In this section, we describe the architecture of the \texttt{Syntriod} algorithm, including template construction, parameter estimation, handling of sparse datasets, and stability-control mechanisms to ensure physically consistent solutions.
\texttt{Syntriod} is a template-based algorithm for estimating Keplerian orbital parameters even from sparse radial velocity observations. The method employs an adaptive architecture that adjusts its operational strategy according to the number of available observations ($N_{\rm obs}$). This section details the algorithm workflow and decision mechanisms.

\subsection{Algorithm Workflow}
\label{sec:algorithm}

{Estimating a unique set of Keplerian orbital parameters from radial-velocity observations requires a sufficient number of independent measurements. In a conventional Keplerian fit, the problem involves six free parameters for SB1 systems $(P, e, \omega, T_0, K, \gamma)$ and seven parameters for SB2 systems when both velocity components are considered. Consequently, a formally constrained solution typically requires at least as many observations as it has parameters.
Unlike conventional time-series approaches, \texttt{Syntriod} performs the inference in the orbital-phase domain by comparing the observations with a precomputed library of Keplerian radial-velocity templates. This template-based framework allows the algorithm to evaluate physically plausible orbital configurations even in sparsely sampled datasets and to provide robust initial estimates even for $N_{\rm obs}=6$ and $5$, where a unique Keplerian solution is formally underconstrained.}

{For datasets with $N_{\rm obs}>5$, \texttt{Syntriod} applies an adaptive period-search strategy and ranks candidate orbital configurations by maximizing the following penalized objective function:}
\begin{equation}
\begin{split}
\ln \mathcal{L}_{\rm eff} =& 
\ln \mathcal{L}_{\rm orb} + W_{\Delta \lambda} \mathcal{L}_{\Delta \lambda}+ W_{dv} C_{dv} \\
& + W_{\nabla} I_{\phi} + W_{\phi} C_{\phi} 
+ \ln \mathcal{P}(e) + \ln \mathcal{P}_{\rm tr}.
\end{split}
\label{eq:eff_l}
\end{equation}
{In this equation, $ \mathcal{L}_{\rm orb}$ represents the base likelihood of the template model. For SB2 systems, the additional terms $L_{\Delta \lambda}$ and $C_{dv}$ account for the consistency between the $RV_1$ and $RV_2$ measurements. The term $I_{\phi}$ gives additional weight to observations located in orbital phases with steep velocity gradients, while $C_{\phi}$ evaluates the phase coverage of the observations. The eccentricity penalty, $\mathcal{P}(e)$, suppresses unrealistically large eccentricity estimates in sparse datasets. The trend penalty, $ \mathcal{P}_{\rm tr}$, suppresses long-period solutions that appear as nearly linear trends over short observational baselines (see Section~\ref{sec:orb_params} for details).}

\begin{figure*}
    \centering
    \includegraphics[width=1\linewidth]{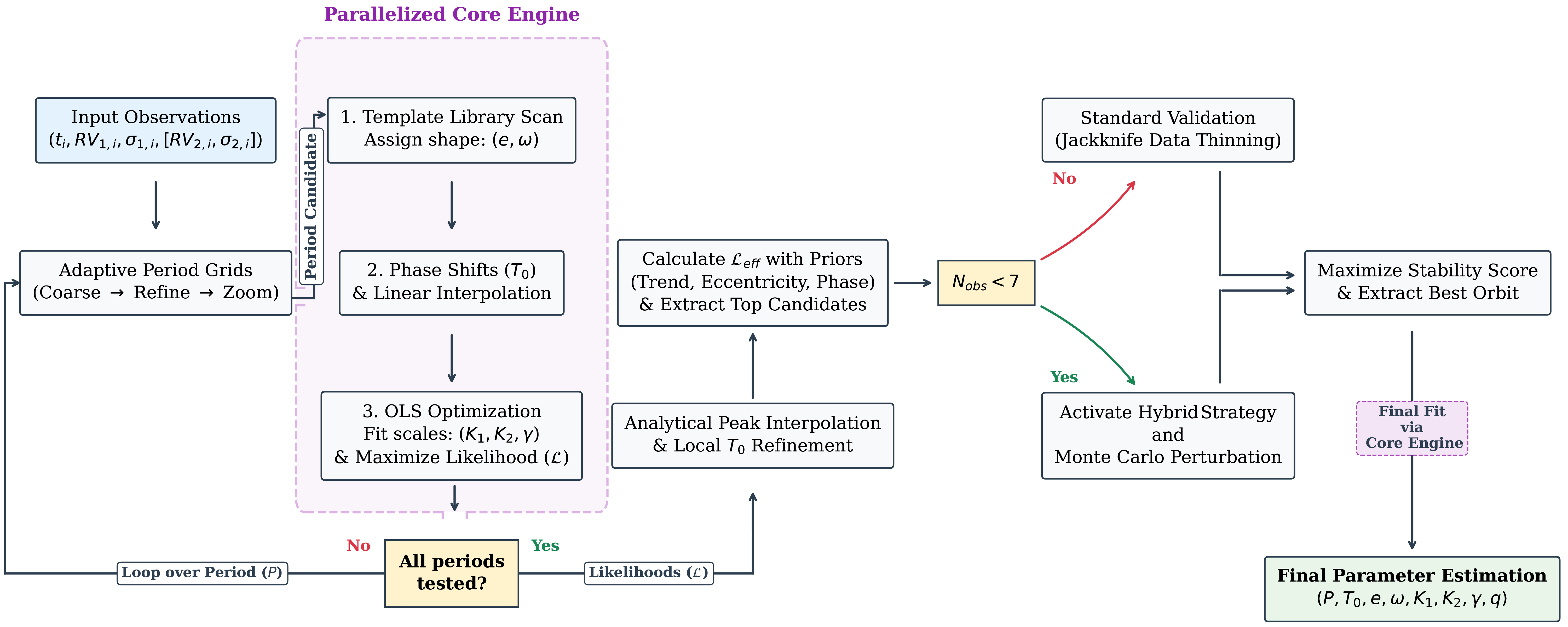}
    \caption{ Schematic workflow of the \texttt{Syntriod} algorithm. The main solution is obtained through adaptive period scanning, phase-domain template matching, and likelihood evaluation. For sparsely sampled datasets ($N_{\mathrm{obs}} < 7$), the auxiliary PSin module and Monte Carlo perturbation branch are activated to reinforce period estimation and solution stability.}
    \label{fig:workflow}
\end{figure*}

{When the number of observations is sufficient for orbital parameter estimation with precomputed templates ($N_{\rm obs}>5$), the algorithm operates in two different regimes depending on the sampling density. Figure~\ref{fig:workflow} summarizes the workflow adopted for orbital parameter estimation when $N_{\rm obs}>5$. For sufficiently sampled datasets ($N_{\rm obs}\ge7$), \texttt{Syntriod} derives the final initialization entirely within the phase-domain template-matching framework and evaluates its stability through a jackknife-like robustness assessment. For sparse datasets ($N_{\rm obs}=6$ or $5$), the robustness assessment relies on Gaussian perturbation tests.  In this regime, the algorithm does not attempt to identify a single unique solution. Instead, it reports the two most likely orbital parameter sets by combining the period estimate from the main period-search approach with an independent estimate from the auxiliary time-domain module, \texttt{PSin}, thereby accounting for aliasing and period ambiguities inherent in sparse sampling.}

{When the number of observations becomes insufficient for a Keplerian solution ($N_{\rm obs}\le4$), \texttt{Syntriod} suppresses the estimation of the shape-dependent orbital parameters ($P$, $e$, and $\omega$) and focuses instead on the linear relations within the radial-velocity measurements. In particular, the mass ratio ($q$) and the systemic velocity ($\gamma$) can be constrained from the linear relation between the component velocities, independently of the orbital period and orbital geometry. \texttt{Syntriod} estimates $q$ and $\gamma$ by applying an ordinary least-squares (OLS) regression \citep{OLS} to $ RV_2 = -q\ RV_1 + \gamma\ (1+q)$ \citep{Wilson}. This approach still provides useful physical information even when only a few RV measurements are available, especially for eclipsing systems with a known orbital period.}


\subsection{Template Library}

Instead of performing direct iterative Keplerian fitting, \texttt{Syntriod} employs a synthetic template-matching strategy based on a library of precomputed theoretical radial-velocity curves. This approach allows simultaneous estimation of the full set of orbital parameters, in contrast to classical periodogram analyses, which typically constrain only the orbital period.

\begin{figure}
    \centering
\includegraphics[width=1\linewidth]{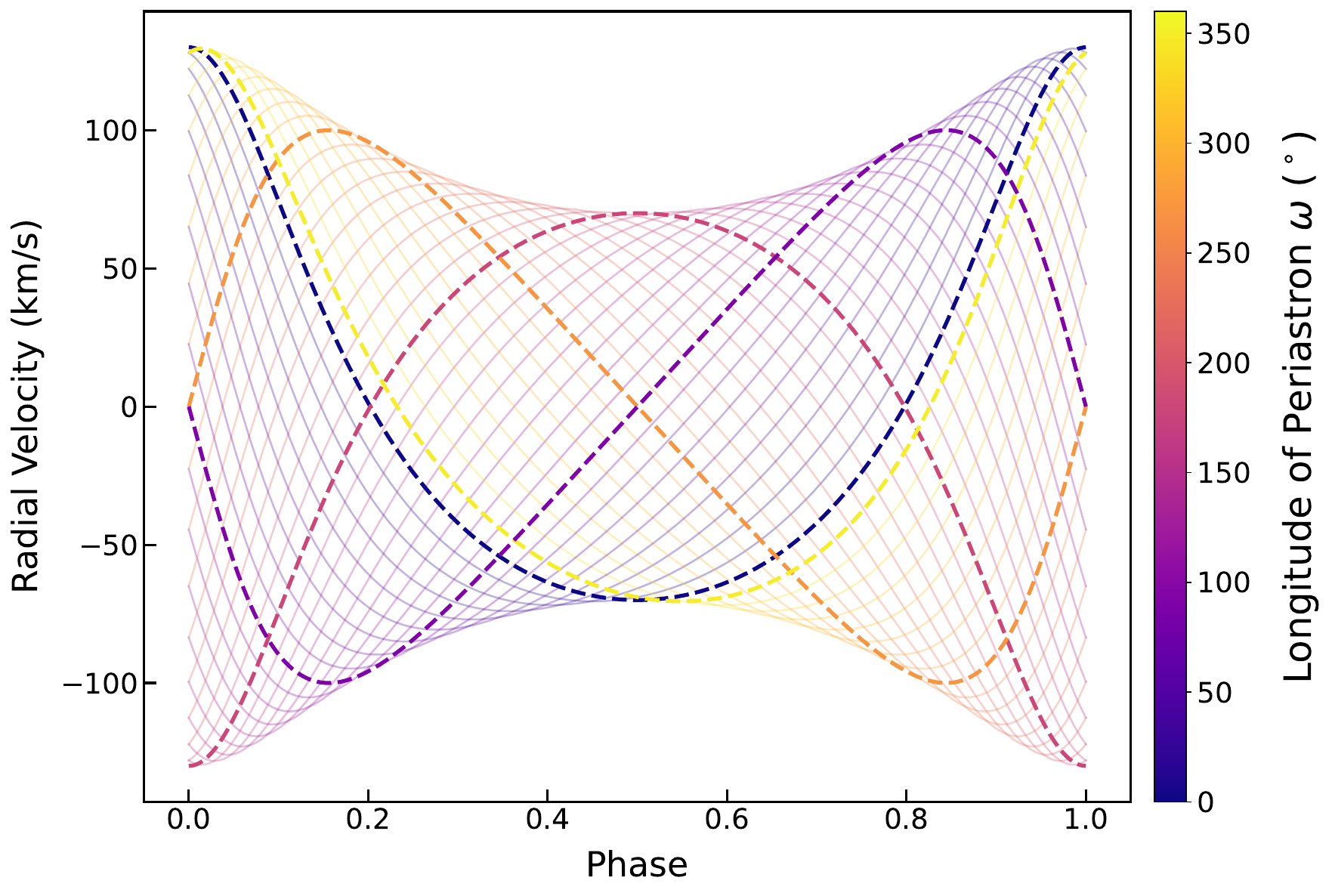}
\caption{Phase-domain visualization of synthetic radial velocity curves from the \texttt{Syntriod} library, fixed at $e=0.3$. The profiles show the variation of the argument of periastron ($\omega$) in $10^\circ$ increments. Dashed lines indicate the specific cases of $\omega = 0^\circ, 90^\circ, 180^\circ$, and $270^\circ$, following the color map.}
\label{fig:templates}
\end{figure}

The library spans a range of orbital geometries defined by eccentricity ($e$) and argument of periastron ($\omega$). {The parameter space is sampled over $e \in [0, 0.8]$ }and $\omega \in [0, 360^\circ)$ with step sizes of $\Delta e=0.1$ and $\Delta\omega=10^\circ$. In the limiting circular case ($e=0$), the argument of periastron becomes degenerate and is therefore restricted to representative values of $0^\circ$,$90^\circ$, $180^\circ$, or $270^\circ$. The resulting template library shown in Figure~\ref{fig:templates} contains {292} distinct Keplerian radial-velocity morphologies. For a fixed eccentricity, variations in $\omega$ modify both the asymmetry and phase structure of the radial-velocity curve. \texttt{Syntriod} exploits these morphological differences to constrain orbital geometry through template matching.  During the construction of the template library, all radial-velocity curves are normalized to a reference orbital period ($P=1\,{\rm d}$) and velocity semi-amplitude ($K=100$ km s$^{-1}$). Each template consists of 1000 phase-sampled representative data points.

{In addition to the standard template library, we provide users with three supplementary template sets tailored for varying precision requirements. These include a moderate-precision set ($\Delta e=0.1$,  $\Delta\omega=5^\circ$; 648 templates) offering slightly higher resolution than the standard grid; an adaptive-resolution set ($\Delta e=0.1$ with $\Delta\omega=10^\circ$ for $e \le 0.3$, refining to $\Delta\omega=5^\circ$ for $e > 0.3$; 473 templates) specifically designed to accurately capture the sharper radial velocity morphologies characteristic of highly eccentric orbits; and a high-precision set ($\Delta e=0.05, \Delta\omega=5^\circ$; 1156 templates) for rigorous parameter constraint. The comparative results of different template sets are in Table~\ref{tab:b1} in Appendix~\ref{B1}.}

\subsection{Period Search Strategy}

To efficiently determine the orbital period while minimizing aliasing risks, \texttt{Syntriod} employs an adaptive period-search strategy rather than a fixed-step scan.
In the first stage, a logarithmic coarse search is performed across the interval $[P_{\rm min}, P_{\rm max}]$. The period grid is constructed in logarithmic space, providing higher resolution at short periods while maintaining computational efficiency at long ones. The number of grid points ($N_{\rm grid}$) is determined dynamically according to both the width of the period interval and the number of observations.

{At the end of the grid search, \texttt{Syntriod} retains the five highest-likelihood period candidates. To avoid nearly identical solutions, the algorithm requires adjacent candidates to satisfy $\Delta P/P > 10^{-3}$. For each candidate period, \texttt{Syntriod} applies a three-point parabolic peak interpolation to the local likelihood distribution. This procedure estimates the exact location of the likelihood maximum and reduces the dependence of the solution on the finite grid resolution.}

{After the algorithm identifies the most likely candidate period ($P_{\rm best}$) through  
the effective likelihood evaluation  (see Eq.~\ref{eq:eff_l}), it performs a two-stage refinement. First, it generates a local search interval spanning $\pm20\%$ around $P_{\rm best}$. It then constructs a second high-resolution interval of $\pm5\%$ around the updated solution. The algorithm adopts the period with the highest final likelihood as the output period estimate.}

{When no prior information about the orbital period is available, imposing a Nyquist-based lower-period bound may appear desirable to reduce aliasing. However, for irregularly sampled radial-velocity observations, a unique Nyquist frequency is generally not defined, and the effective alias structure depends on the spectral window of the observing times \citep{nyquist, Eyer99}. Therefore, \texttt{Syntriod} does not enforce a hard lower-period limit based on the median cadence. Instead, users may optionally activate an empirical lower bound, such as $P_{\rm min}=0.2\,\Delta T_{\rm median}$, to reduce the explored parameter space \citep[e.g.,][]{Sanders06}. In this work, we do not apply such a constraint and instead allow the period search to span the full predefined interval.
}

\subsubsection{The PSin Module}

In sparse datasets, the detailed morphology of the radial velocity curve becomes difficult to resolve, although the fundamental periodicity may still be detectable. The PSin module is {a periodogram} designed to identify the dominant harmonic component of the signal independently of orbital geometry.
{The method operates in the time-domain and focuses on how the observations vary with time. At this stage, the goal is not to reconstruct the full orbit. Instead, PSin aims to identify a reliable period range that \texttt{Syntriod} can later explore in detail using its template library framework.}

{\texttt{PSin} builds on the mathematical framework of the GLS periodogram \citep{GLS}, acting as a maximum-likelihood estimator in the phase-folded time domain. Whereas the standard GLS uses a floating-mean model, \texttt{PSin} extends this approach by incorporating first-order harmonic terms that effectively capture the asymmetric radial velocity curves characteristic of eccentric orbits ($e > 0.3$). During the period search, PSin fits the following extended trigonometric model to the radial-velocity measurements:}

\begin{equation}
\begin{split}
RV(t) &= C + A_1 \sin\left(\frac{2\pi t}{P}\right)
      + B_1 \cos\left(\frac{2\pi t}{P}\right) \\
      &\quad + A_2 \sin\left(\frac{4\pi t}{P}\right)
      + B_2 \cos\left(\frac{4\pi t}{P}\right)
\end{split}
\end{equation}

{In this equation, $C$ represents the systemic velocity ($\gamma$), $A_1$ and $B_1$ describe the fundamental frequency, and $A_2$ and $B_2$ describe the first harmonic, which captures deviations from a purely sinusoidal orbit. This model does not require an iterative optimization procedure. The vector of unknown coefficients ($\mathbf{\beta}$) is determined analytically in a single step using Weighted Linear Least Squares (WLS) regression \citep{OLS}:}

\begin{equation}
\mathbf{\beta}
=
({X}^T\ {W}\ {X})^{-1}
{X}^T\ {W}\ {y}.
\end{equation}
{Here, ${X}$ denotes the design matrix, ${W}$ is the weight matrix derived from the observational uncertainties, and ${y}$ is the radial-velocity vector.}
{To improve computational efficiency, PSin uses a two-stage optimization grid. It first performs a coarse scan over a broad period range and then applies a high-resolution search around the strongest candidate peak.}

{The classical Lomb--Scargle periodogram estimates the spectral power density of the data and often assumes a zero-mean signal for computational simplicity. In sparse and irregularly sampled observations, this assumption can lead to phase aliasing and incorrect period estimates. PSin reduces the effects of this limitation by solving for the systemic velocity simultaneously through the constant term ($C$) in the design matrix. In addition, its harmonic structure follows the logic of a truncated Fourier series, allowing the method to describe mildly asymmetric orbital signals. These features make PSin substantially more stable than standard LS and GLS when only a small number of observations are available. The resulting candidate period ($P_{\mathrm{PSin}}$) is then used to guide \texttt{Syntriod}'s subsequent template-based refinement within a $\pm10\%$ window.}

\subsection{Estimation of Orbital Parameters}\label{sec:orb_params}

{ 
For a given period $P$, \texttt{Syntriod} estimates the remaining Keplerian parameters through a deterministic sequence of steps. The primary component of the effective likelihood definition given in Equation \ref{eq:eff_l}, $\mathcal{L}_{\rm orb}$, relies on comparisons between observations and precomputed templates, and its computational steps are as follows. As the first step, observation times are phase-folded using $\phi = (t-T_0)/P \pmod 1$. Since the true time of periastron passage ($T_0$) is initially unknown, the templates must be systematically shifted along the orbital-phase axis to identify the optimal data-model alignment. Furthermore, because $T_0$ exhibits a strong covariance (partial degeneracy) with the argument of periastron ($\omega$), this phase domain must be sampled with high resolution to prevent parameter mischaracterization. The algorithm therefore scans the $T_0$ parameter using an adaptive grid with a density that scales approximately as $1000/N_{\rm obs}$.}


Since the template library contains normalized velocity amplitudes, to recover the physical amplitude of the primary component, OLS regression is performed between the observed velocities ($RV_1$) and the template model:
\begin{equation}
\chi^2_{OLS} = \sum_i (RV_{1,i} - (S \cdot \mathcal{X}_i + c))^2
\end{equation}  
Here, $\mathcal{X}i$ denotes the template value at the corresponding orbital phase. The regression slope $S$ yields the semi-amplitude $K_1 = |100 \cdot S|$ where the factor of 100 accounts for the normalization of the template. The intercept $c$ corresponds to the systemic velocity $\gamma$, which can be written analytically as 
\begin{equation}
\gamma = \frac{\sum RV_{1,i} - S \sum \mathcal{X}_i}{N_{obs}},
\end{equation} where $N_{obs}$ is the number of observations. This analytical solution allows the velocity-scaling parameters to be determined without iterative optimization. 

Once the parameters $(P, T_0, K_1, \gamma)$ are determined, the remaining orbital parameters $(e,\omega)$ are constrained through template matching.
The quality of each template solution is evaluated using the Gaussian likelihood function \citep{GL}
\begin{equation}
\begin{split}
    \ln \mathcal{L}_{\rm orb} = -\frac{1}{2} &\Biggl[ \sum_{i} \frac{(RV_{1,i} - RV_{mdl,i})^2}{\sigma^2} \\
    &\quad + N \ln(2\pi\sigma^2) \Biggr]
\end{split}
\label{eq:likelihood_orb}
\end{equation}
where $RV_{mdl}$ denotes the scaled template model ({$RV_{mdl} = \gamma + K_1  \mathcal{X}_{tpl}$}). The likelihood formulation incorporates observational uncertainties directly into the model evaluation and provides a consistent probabilistic metric for comparing different template solutions.

\subsubsection{The Differential Velocity Approach in SB2 Systems}

The second and third terms of the $\mathcal{L}_{\rm eff}$ (Equation \ref{eq:eff_l}) come into play for SB2 systems; the former computes the likelihood by accounting for the radial velocity differences between the two components. In the analysis of double-lined spectroscopic binaries (SB2), \texttt{Syntriod} treats the radial velocity curves of both components ($RV_1$ and $RV_2$) as a coupled dynamical system governed by identical orbital mechanics. For a given candidate orbital period and template morphology ($\mathcal{X}_{\mathrm{tpl}}$), the algorithm simultaneously derives the systemic velocity ($\gamma$) and velocity semi-amplitudes ($K_1$, $K_2$) by reducing the OLS problem to a compact $3\times3$ linear system. The formulation explicitly minimizes the total residual sum of squares:
\begin{equation}
RSS_{\mathrm{total}} =
\sum_{j=1}^{2} \sum_{i=1}^{N}
\left(
RV_{j,i} -
(\gamma \pm K_j \mathcal{X}_i)
\right)^2 ,
\end{equation}
where $\mathcal{X}_i$ represents the normalized template value at the corresponding orbital phase.

Solving the resulting $3\times3$ system analytically provides a computationally efficient alternative to iterative fitting while ensuring dynamical consistency between the two velocity curves. To further suppress degenerate solutions (e.g., harmonic configurations producing parallel velocity curves), \texttt{Syntriod} incorporates the differential velocity formalism of $\Delta v = v_1 - v_2$ as an additional regularization term \citep{Hilditch}.
The instantaneous velocity difference predicted by the model is 
\begin{equation}
\Delta v_{\mathrm{mdl}} = (K_1 + K_2)\,\mathcal{X}(\phi).
\end{equation}

The likelihood function, therefore, {defines the differential velocity constraint}:
\begin{equation}
\begin{split}
    W_{\Delta v} \ln \mathcal{L}_{\Delta v} = -\frac{W_{\Delta v}}{2} &\Biggl[ \sum_{i} \frac{(\Delta v_{obs,i} - \Delta v_{mdl,i})^2}{\sigma_1^2 +\sigma_2^2} \\
    &\quad + N \ln(2\pi(\sigma_1^2 +\sigma_2^2)) \Biggr]
\end{split}
\label{eq:likelihood}
\end{equation}
Here $\mathcal{L}_{\Delta v}$ measures the agreement between the observed and modeled velocity differences, and $W_{\Delta v}$ controls the relative weight of the regularization term, {and is set to be 0.3.}

The velocity-separation score, the third term in Equation~\ref{eq:eff_l}, $W_{dv} C_{dv}$, also applies exclusively to SB2 systems. It quantifies the fraction of the orbital phase in which the velocity difference between the two components exceeds a predetermined threshold. Mathematically, this score is defined as:
$$C_{dv} = \int_{0}^{1} \Theta \left( |\Delta v(\phi)| - f_{\mathrm{th}} \Delta v_{\max} \right) d\phi$$
where $\Delta v(\phi)$ is the modeled instantaneous velocity difference, $\Delta v_{\max}$ is the global maximum of this separation, $f_{\mathrm{th}}$ is the fractional threshold constraint (set to 0.6 in our grid), and $\Theta(x)$ represents the Heaviside step function. By penalizing models where the integration yields a low $C_{dv}$ value, the algorithm effectively suppresses heavily blended or unresolved alias solutions.

\subsubsection{{Auxiliary Control Terms}}

During likelihood maximization, the algorithm applies auxiliary control terms to suppress physically implausible or poorly constrained results.

To quantify the quality of orbital phase sampling, \texttt{Syntriod} additionally evaluates the gradient of the normalized template shape $\mathcal{X}(\phi)$, producing an information map defined by $ \mathcal{X}(\phi)=\left|
{d\mathcal{X}}/{d\phi}
\right|$. Observations obtained near phases with steep velocity gradients (for example, around periastron passage) provide stronger constraints on orbital geometry. \texttt{Syntriod} therefore projects the observational cadence onto this gradient map to compute a phase-information score:
\begin{equation}
I_{\phi}
=
\frac{1}{N}
\sum_{i=1}^{N}
\left|
\frac{d\mathcal{X}}{d\phi}
\right|_{\phi_i}.
\end{equation}
A high value of $I_{\phi}$ indicates that the dataset samples dynamically informative orbital phases.

{One of the Bayesian penalty terms is the Phase Coverage Score ($C_{\phi}$), which penalizes solutions that leave large gaps in orbital phase coverage ($\Delta\phi_{max} \approx 1$). This approach helps reject misleading models that rely on poorly sampled orbital configurations. Mathematically, these metrics are defined as:}
\begin{equation}
W_{\phi} C_{\phi} = W_{\phi}(1 - \Delta\phi_{max}),
\end{equation}
{where $\Delta\phi_{max}$ is the maximum distance between the consecutive phases and the weight $W_{\phi}= 1.5$.}

{A second penalty term addresses the tendency of sparse data sets to favor artificially large eccentricities when the period estimate is incorrect. To reduce this effect, \texttt{Syntriod} applies an eccentricity penalty that scales inversely with the number of observations:
\begin{equation}
\ln \mathcal{P}(e) = - \left( \frac{N_{0}}{\max(N_{obs}, 1)} \right) \left( \frac{e}{e_0} \right)^2.
\end{equation}
In this expression, $N_{0}$ defines a reference observation count and has a value of 8. The parameter $e_0$ defines the characteristic eccentricity scale and is set to 0.4. We choose $e_0=0.4$ because orbital asymmetries become clearly visible above this eccentricity. As a result, the penalty remains weak for well-sampled systems but increasingly suppresses unrealistically high eccentricity solutions when only a small number of observations are available.}

{Long-period systems may appear approximately linear over short observational baselines. To prevent the algorithm from interpreting such trends as genuine orbital solutions, \texttt{Syntriod} compares the likelihood of each proposed orbital model ($\ln\mathcal{L}_{orb}$) with the likelihood of a simple linear trend model ($\ln\mathcal{L}_{tr}$):
\begin{equation}
    \ln \mathcal{P}_{tr} = \begin{cases} -W_{tr} \left( 1 - \frac{\Delta}{\Delta_{0}} \right) & \text{if } \Delta < \Delta_{0} \\ 0 & \text{if } \Delta \ge \Delta_{0} \end{cases}
    \end{equation}
where $\Delta = \ln \mathcal{L}_{orb} - \ln \mathcal{L}_{tr}$.
The parameter $W_{tr}$ controls the penalty strength and is set to 2. The parameter $\Delta_{0}$ defines the minimum likelihood improvement required for an orbital solution and is set to 3.}

\subsection{{Robustness Assessments}}

To assess the reliability of orbital solutions under varying sampling conditions, \texttt{Syntriod} adopts a dynamic validation strategy that depends on the number of available observations ($N_{\mathrm{obs}}$). {For datasets with sufficient phase coverage (i.e., $N_{\mathrm{obs}} \geq 7$), \texttt{Syntriod} evaluates solution stability using a randomized data-thinning procedure inspired by the Jackknife method \citep{Jackknife}. It is important to clarify that this procedure serves strictly as a robustness and leverage test, rather than a formal jackknife uncertainty estimator. Because \texttt{Syntriod} is fundamentally designed as an initial parameter estimation stage to find the global minimum basin, it does not aim to output formal parameter uncertainties. During this process, a randomly selected fraction (typically $15\%-25\%$) of the observations is temporarily removed, and the orbital solution is recomputed iteratively. If the recovered orbital parameters remain stable across these trials, the solution is considered robust and representative of the global structure of the dataset rather than leveraged by a single outlier or an alias configuration.}

When the number of observations falls below seven ($N_{\mathrm{obs}} < 7$), removing even a single data point may critically disrupt orbital phase coverage. In this regime, \texttt{Syntriod} replaces the data-thinning strategy with a Monte Carlo perturbation approach \citep{numerical}. Synthetic realizations of the dataset are generated by injecting Gaussian noise {(scaled by 0.1)} into the observed radial velocities:
\begin{equation}
RV_{\mathrm{test},i} =
RV_{\mathrm{obs},i} +
\mathcal{N}(0,\sigma_{\mathrm{perturb}})
\end{equation}
This procedure evaluates the sensitivity of the orbital solution to measurement uncertainties while preserving the dataset's temporal sampling. If the recovered period remains stable across perturbation realizations, the detected signal is considered statistically robust.

\subsection{Computational Complexity and Runtime Performance}

The computational cost of \texttt{Syntriod} is determined primarily by the size of the period grid, the number of template evaluations, and the number of observations in the dataset. Because the algorithm relies on deterministic template matching rather than iterative global optimization, its runtime remains predictable and scales approximately linearly with the number of trial periods and templates.

For each candidate period, the algorithm performs the following operations: (i) phase folding of the observations, (ii) scanning of the $T_0$ grid, (iii) linear estimation of the scaling parameters ($K_1$, $\gamma$), and (iv) template evaluation over the $(e,\omega)$ library. If the number of candidate periods is denoted by $N_P$, the number of phase-shift trials by $N_{T_0}$, the number of templates by $N_{\mathrm{tpl}}$, and the number of observations by $N_{\mathrm{obs}}$, the dominant cost can be approximated as
\begin{equation}
\mathcal{O}\!\left(
N_P \, N_{T_0} \, N_{\mathrm{tpl}} \, N_{\mathrm{obs}}
\right).
\end{equation}

Since the template library is fixed in size and the number of observations in sparse-data applications is typically small, the effective runtime is dominated by the adaptive period grid. The logarithmic coarse search and local refinement strategy substantially reduces the total number of period evaluations compared to a uniform dense scan over the full parameter space.

On a standard computer (e.g., 2.3 GHz processor with 14 cores and 16 GB RAM), the computation time of \texttt{Syntriod} is approximately 5 seconds for datasets where $N_{\mathrm{obs}} > 10$, evaluating around $20 \times 10^6$ iterations. This high computational throughput is achieved through a highly efficient architecture powered by Just-In-Time (JIT) compilation \citep{lam2015}. Conversely, for cases where $N_{\mathrm{obs}} \le 10$, the resolution of the period grid is increased inversely proportionally to the decrease in the number of data points. Consequently, the computational load scales up to approximately $60 \times 10^6$ iterations for a dataset with $N_{\mathrm{obs}} = 5$. Due to the enhanced scanning resolution, the runtime for systems with a {few} observations can reach $\sim$10 seconds. Furthermore, upon the very first execution of the routine, a one-time initial overhead of approximately 5 seconds is incurred to compute the gradient map of the template library.

In sparse-data regimes ($N_{\mathrm{obs}} < 7$), the activation of the auxiliary PSin module introduces an additional computational step. However, because PSin is solved analytically through weighted linear least squares, its cost remains low compared to the full template-matching stage. As a result, the hybrid architecture preserves computational efficiency while improving robustness in difficult sampling regimes.

In practical applications, \texttt{Syntriod} is intended not only as a standalone orbital parameter estimator but also as a low-cost pre-solver for more computationally intensive orbit-fitting pipelines. By constraining the parameter space near the physically relevant solution, the algorithm can substantially reduce the initialization cost for subsequent inference methods.

\section{Validation Tests and Results with Synthetic Orbits}
\label{sec:test_results}

We assess the accuracy and precision of \texttt{Syntriod} by obtaining orbital parameter predictions via sampling random data points from synthetic radial velocity curves generated from the theoretical Keplerian orbits.
{To facilitate interpretation of the recovery statistics, we divide the simulations into four observational regimes according to the number of radial-velocity measurements. The first regime, $N_{obs} \geq 10$, represents the recommended sampling domain \citep[e.g.,][]{rvfit} and is broadly consistent with commonly adopted observational guidelines for reliable spectroscopic-binary orbit determination. The second regime, $N_{obs} = 9$ and $8$, corresponds to a moderately sampled domain where orbital initialization remains generally reliable.
The third regime, $N_{obs}=6$, represents the formal dimensional limit of the Keplerian SB1 problem. 
The fourth regime, $N_{obs}=5$, is treated as a stress-test domain in which the problem formally becomes underconstrained with respect to a unique solution. }

\subsection{Keplerian-Based Synthetic Data}
\label{sentetik}

To test the prediction reliability of \texttt{Syntriod}, we generate 10\,000 synthetic radial velocity curves using a Python routine based on the theoretical Keplerian orbit. This sample provides a statistically robust and realistic set of mock systems. To isolate the intrinsic solution accuracy across a large statistical sample, we avoid injecting observational noise into the synthetic RV curves {(See Section~\ref{A1} in Appendix for results with noise-injected synthetic data.)}.

\begin{figure}[h]
    \centering
    \includegraphics[width=\linewidth]{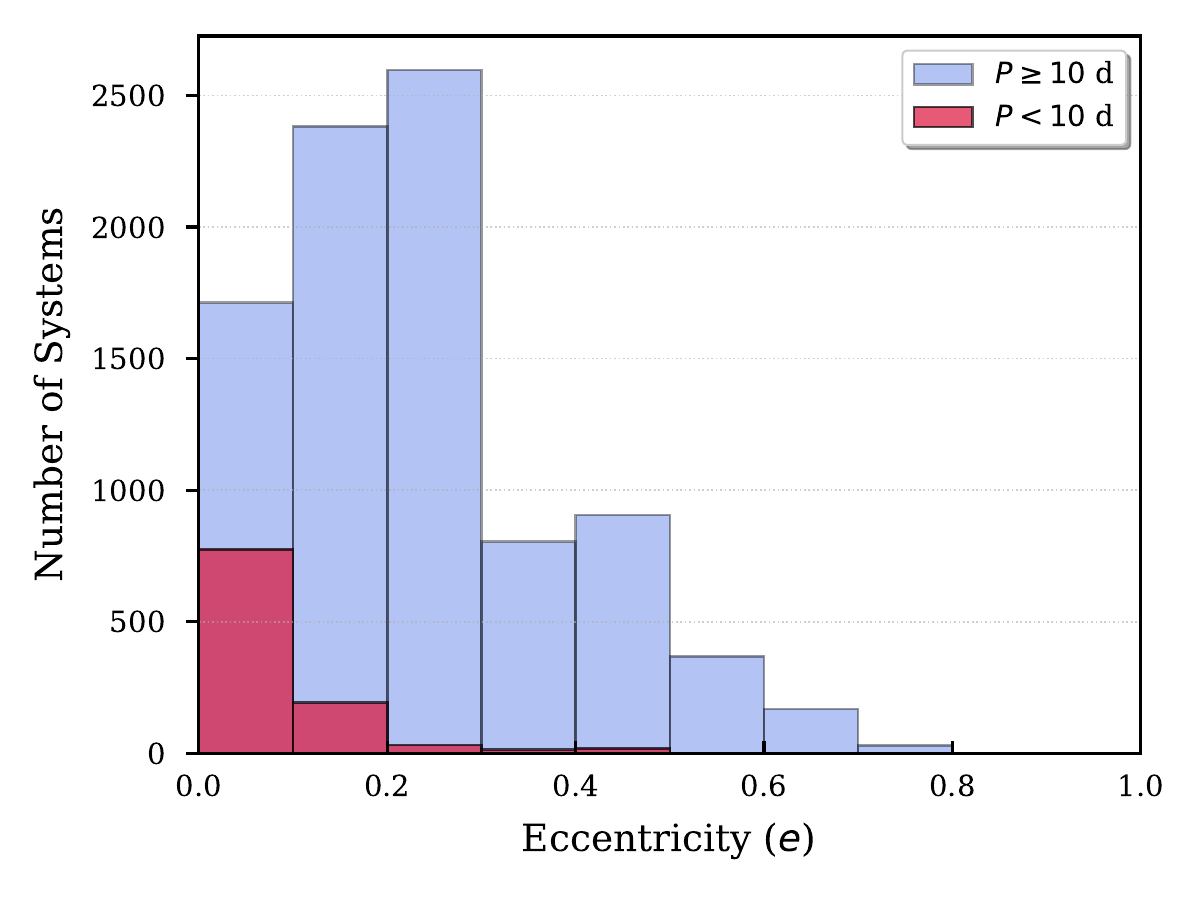}
    \caption{Eccentricity distribution as a function of orbital period ($P$) for the synthetic radial velocity data generated to test the \texttt{Syntriod} algorithm.}
    \label{fig:e_dist}
\end{figure}

The orbital periods of the synthetic orbits are equally distributed within the range $P \in [0.1, 100]$ days (See Section ~\ref{sec:real_sb} for examples reaching beyond these limits).  
This range encompasses the majority of spectroscopic detections while ensuring a statistically representative period distribution and sufficient temporal coverage between data points. Figure~\ref{fig:e_dist} compares the eccentricity distributions of the synthetic orbits for short- and longer-period regimes of the mock data.
The {eccentricities cover the interval {$e \in [0, 0.8]$}}, following period-dependent probability distributions. 
Short-period systems ($P < 10^d$) favor low eccentricities, with a decreasing probability toward higher $e$ within the range $[0, 0.2]$. Longer-period systems ($P > 10^d$) span the full eccentricity range and show a broad maximum near $e \approx 0.2 - 0.3$. These distributions reflect the general tendency of the real observed systems \citep{Halbwachs, Raghavan, tokovinin, Pourbaix}. 

We create mock observed RV curves by randomly sampling data points from the synthetic orbits. To evaluate the sensitivity of \texttt{Syntriod} to sampling density, {we generate non-nested datasets for the optimal sampling regime with $N_{obs}=10$, the moderately sampled regime with $N_{obs}=8$, the formal solution limit with $N_{obs}=6$, and the stress test regime with $N_{obs}=5$.}  In all cases, we randomly sample the observation times to ensure unbiased phase coverage.
{For each synthetic binary star system, the observational baseline ($T_{\rm span}$) was drawn randomly from a log-normal distribution centered around a median of 10,000 days to reflect realistic survey strategies and data sparsity driven by telescope time constraints. This sampling technique allows the $T_{\rm span}/P$ ratios to range from 0.1 to 1000, peaking at $\sim 20$ (see Figure~\ref{fig:T_span_Phi}), thereby establishing a demanding test configuration to measure the algorithm's robustness to incomplete phase coverage.}

\subsection{Performance of Period Estimation on Synthetic Data}
\label{period estimation}

The orbital period $P$ constitutes the fundamental parameter in the solution of spectroscopic binary orbits because it defines the time scale of the Keplerian motion and determines the phase of all observations. The radial velocity signal depends explicitly on orbital phase, $\phi = (t - T_0)/P$, and therefore an incorrect estimation of $P$ propagates directly into the inferred values of $e$, $K$, $\omega$, and $T_0$ \citep{lucy1971}. Classical treatments of spectroscopic binaries emphasize that the period must be established before a reliable nonlinear orbital fit \citep[e.g.,][]{Hilditch, Pourbaix}. 
In general practice, period determination is treated as a signal-detection problem in unevenly sampled data, commonly addressed using Lomb–Scargle–type methods \citep{GLS,lomb}.

We first focus on period-estimation accuracy, comparing \texttt{Syntriod} predictions with those from the generalized Lomb–Scargle (LS) implementation in the \texttt{astropy} library. We define a tolerance based on the relative difference between the recovered and true periods, i.e. \begin{equation}
    \frac{|\Delta P|}{P} = \frac{|P_{\mathrm{est}} - P_{\mathrm{real}}|}{P_{\mathrm{real}}}.
\end{equation}
We characterize a 
\textit{successful estimate} when this tolerance remains below 10\% and  a \textit{precise estimate} when it falls below 1\%. {Note that since the \texttt{Syntriod} algorithm produces the two most possible orbital parameter sets as initial estimates for {$N_{obs}<7$}, we calculate accuracy ratios considering either of these two solutions is within the range.\footnote{Secondary predictions from the PSin module increase the success rate by 5–10\% for cases with $N_{\rm obs}=6$ and $5$, respectively.}}  

\begin{figure}[h]
    \centering
    \includegraphics[width=1\linewidth]{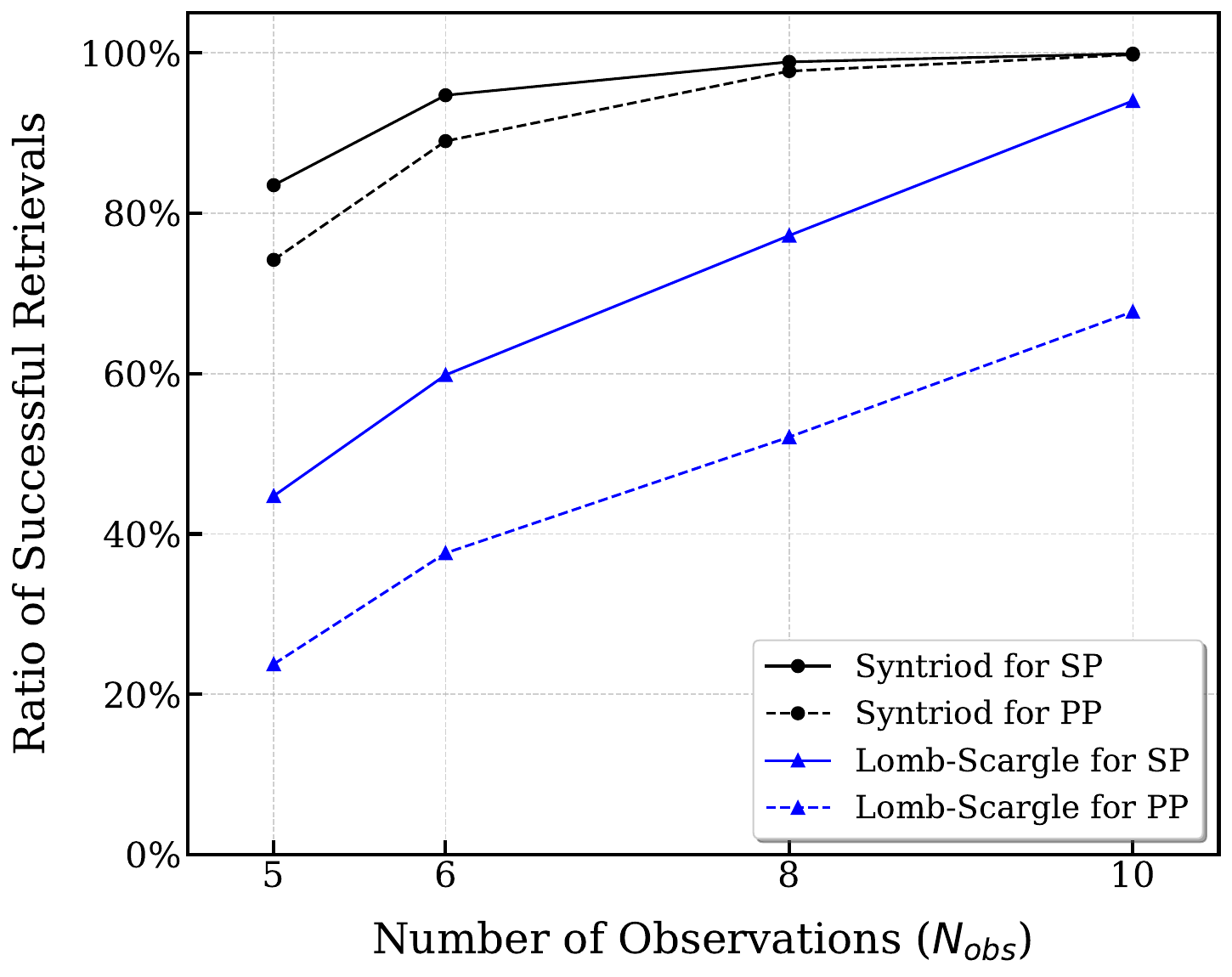}
    \caption{The distribution of successful (lines) and precise (dashed lines) period prediction rates of \texttt{Syntriod} (black circles) and Lomb–Scargle (blue triangles) under varying numbers of data points drawn from 10\,000 synthetic systems. The successful predictions have a tolerance of less than 10\%, and the precise predictions have a tolerance of less than  1\%. }
    \label{fig:Synt_vs_LS}
\end{figure}

Figure~\ref{fig:Synt_vs_LS} presents the fraction of successful and precise period estimates from both \texttt{Syntriod} and LS as a function of the number of observations drawn from the 10\,000 synthetic datasets. 
In the optimal sampling {regime} with $N_{obs}=10$, both methods achieve high successful estimate rates (i.e., tolerance $\leq$ 10\%), reaching {99.90\%} for \texttt{Syntriod} and {94.01\%} for LS. However, when considering precise estimates (tolerance $\leq$ 1\%), \texttt{Syntriod} attains a rate of {99.78\%}, whereas LS reaches only {67.72\%}.

As $N_{obs}$ decreases, the performance gap between the two methods becomes more pronounced. For instance, at {the formal solution limit with} 
$N_{obs}=6$, the successful estimate rate remains as high as {94.71\%} for \texttt{Syntriod}, whereas it drops to about {61.91\%} for LS. The results indicate that achieving a success rate above 90\% requires about 
$N_{obs}=10$ for LS, while \texttt{Syntriod} reaches the same threshold  at 
$N_{obs}=6$.
A similar trend is observed for precise estimates. The precise-estimate rate of LS decreases rapidly with decreasing 
$N_{obs}$, whereas for \texttt{Syntriod}, the successful and precise rates remain close to each other for 
$N_{obs}\geq5$. This finding demonstrates that \texttt{Syntriod} not only requires fewer observations to maintain high success rates but also preserves its precision significantly better than LS.
Even when the number of observations falls below the number of free orbital parameters (i.e., {$N_{\rm obs}=5$}), the template–matching approach of \texttt{Syntriod} yields success rates above 80\% while LS is suffering from aliasing effects in these sparse regimes, which reduces the fraction of correct period identifications.

\begin{figure*} 
    \centering
\includegraphics[width=\textwidth]{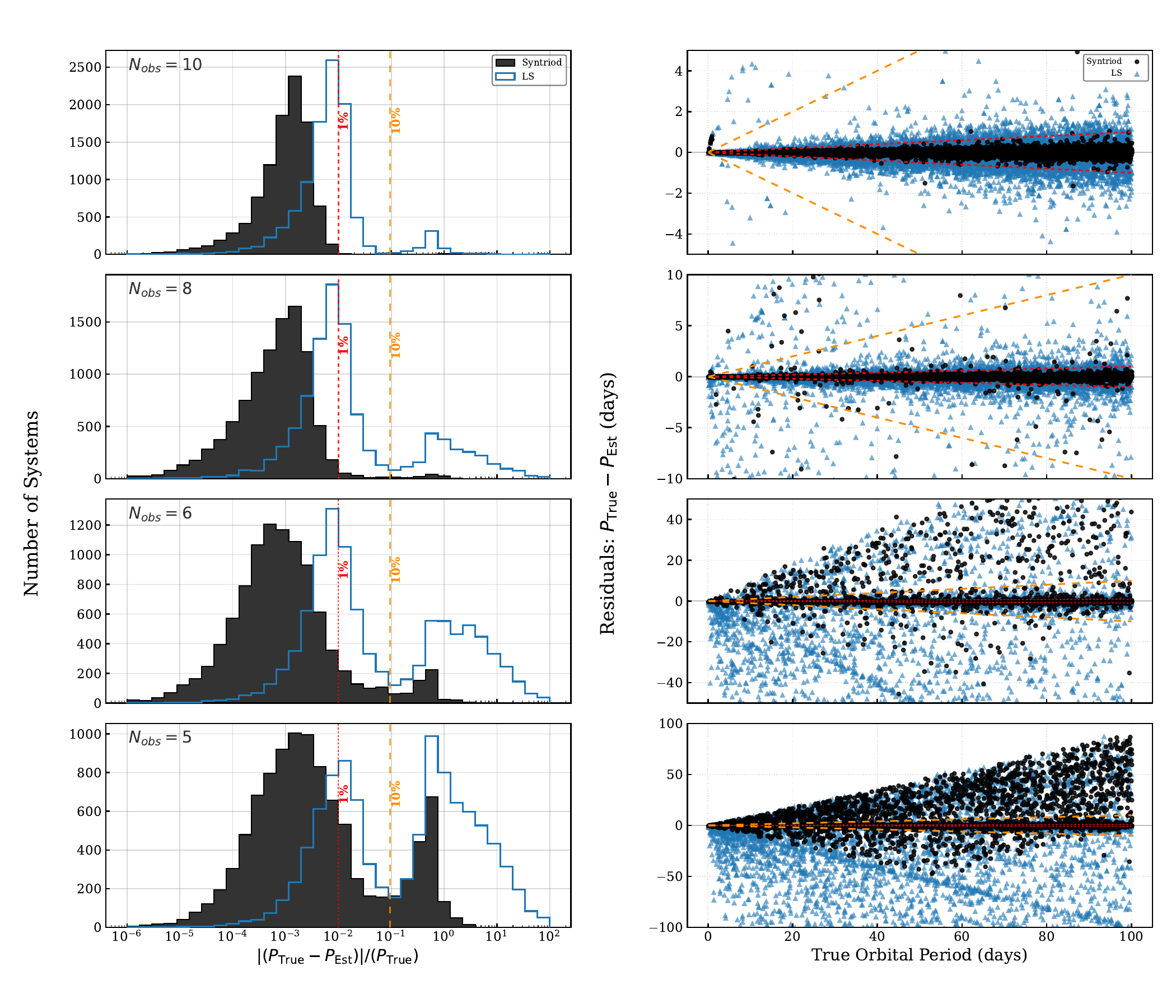}
    \caption {(Left panels) The distribution of the relative period difference, $(P_{\rm est}-P_{\rm true})/P_{\rm true}$, obtained with \texttt{Syntriod} (black) and LS (blue) for  $N_{\rm obs}=10, 8, 6,$ and $5$ from top to bottom. (Right panels) The dependence of the period residuals on the true period.}
    \label{fig:8-panels}
\end{figure*}

Figure \ref{fig:8-panels} shows the distribution of the relative period difference, $(P_{\rm est}-P_{\rm true})/P_{\rm true}$, obtained with the two methods for different numbers of radial-velocity observations. The panels are arranged by decreasing sampling density, with $N_{\rm obs} = 10, 8, 6,$ and $5$ from top to bottom. The left panels show histograms of relative period differences, while the right panels show the relationship between the estimated period and its deviation from the true period.

The relative error distributions exhibit a characteristic bimodal morphology (see Figure~\ref{fig:8-panels}), with a dominant peak at small relative differences corresponding to correct solutions and a secondary peak from incorrect period identifications. Hartigan’s Dip test \citep{dip} confirms statistically significant bimodality for the Lomb–Scargle (LS) solutions in all panels ($p \leq 0.01$). In contrast, the \texttt{Syntriod} results do not show significant bimodality for most sampling regimes ($p>0.99$ for $N_{\rm obs}=10$, $8$,  $6$, and $5$, indicating that the alias-driven population is largely suppressed.

For \texttt{Syntriod}, the secondary mode is negligible for $N_{\rm obs}=10$ and $N_{\rm obs}=8$, solutions are tightly clustered around the true period. A visible population of incorrect solutions appears only at {the formal solution limit with} $N_{\rm obs}=6$ and becomes more pronounced at $N_{\rm obs}=5$. Even in this sparse regime, the dominant population remains confined below the $10\%$ tolerance limit, and for $N_{\rm obs}\geq6$ most solutions lie within the $1\%$ precision threshold. The global median relative differences in the \texttt{Syntriod} distributions range from {$0.0011$ to $0.0024$}, corresponding to a typical accuracy of roughly $10^{-3}$ in period recovery. This indicates that \texttt{Syntriod} maintains both a high correct-recovery fraction and high precision in the recovered periods, with performance degradation manifesting primarily as a gradual broadening of the uncertainty distribution.

In contrast, LS already exhibits a clearly separated secondary peak at $N_{\rm obs}=10$, demonstrating that alias-driven misidentifications occur even under relatively well-sampled conditions. As $N_{\rm obs}$ decreases, the amplitude of this secondary mode increases substantially. The median relative differences of the LS distributions span a much wider range, from {$0.0070$ to $0.3250$}, reflecting the increasing influence of incorrect alias solutions. In this case, the degradation is dominated not by a smooth loss of precision but by discrete transitions to alternative periodicities imposed by the sampling window.

The right panels of Figure \ref{fig:8-panels} illustrate the dependence of the period residuals on the true period. For visual guidance, the curves corresponding to the $1\%$ and $10\%$ relative tolerance limits are shown, while the vertical axis limits are adjusted in each panel to optimize visibility. No systematic clustering of incorrect solutions is observed at specific periods. However, the LS results reveal clear linear structures in the $P$--$\Delta P$ plane for the incorrect solutions. These patterns indicate that the recovered periods differ from the true value by approximately integer multiples of the period, consistent with aliasing, in which the true periodicity is confused with its harmonics due to the sampling window. Not all incorrect solutions lie close to harmonic relations; however, a noticeable level of general scatter is also present across the parameter space.

\subsection{{{Dependence of Period Recovery on Orbital and Sampling Parameters}}}\label{sec:3.3}

{As the next step of the validation tests, we examine the period recovery performance of \texttt{Syntriod} and the LS algorithm as a function of the orbital parameters $P$ and $e$, and the observational quantities $T_{\mathrm{span}}$ and $\Delta \phi_{\max}$. To isolate the specific effects of data sparsity, we analyze the simulation results separately for $N_{\mathrm{obs}} = 10, 8, 6$, and $5$.}

\begin{figure*}
    \centering
    \includegraphics[width=1\linewidth]{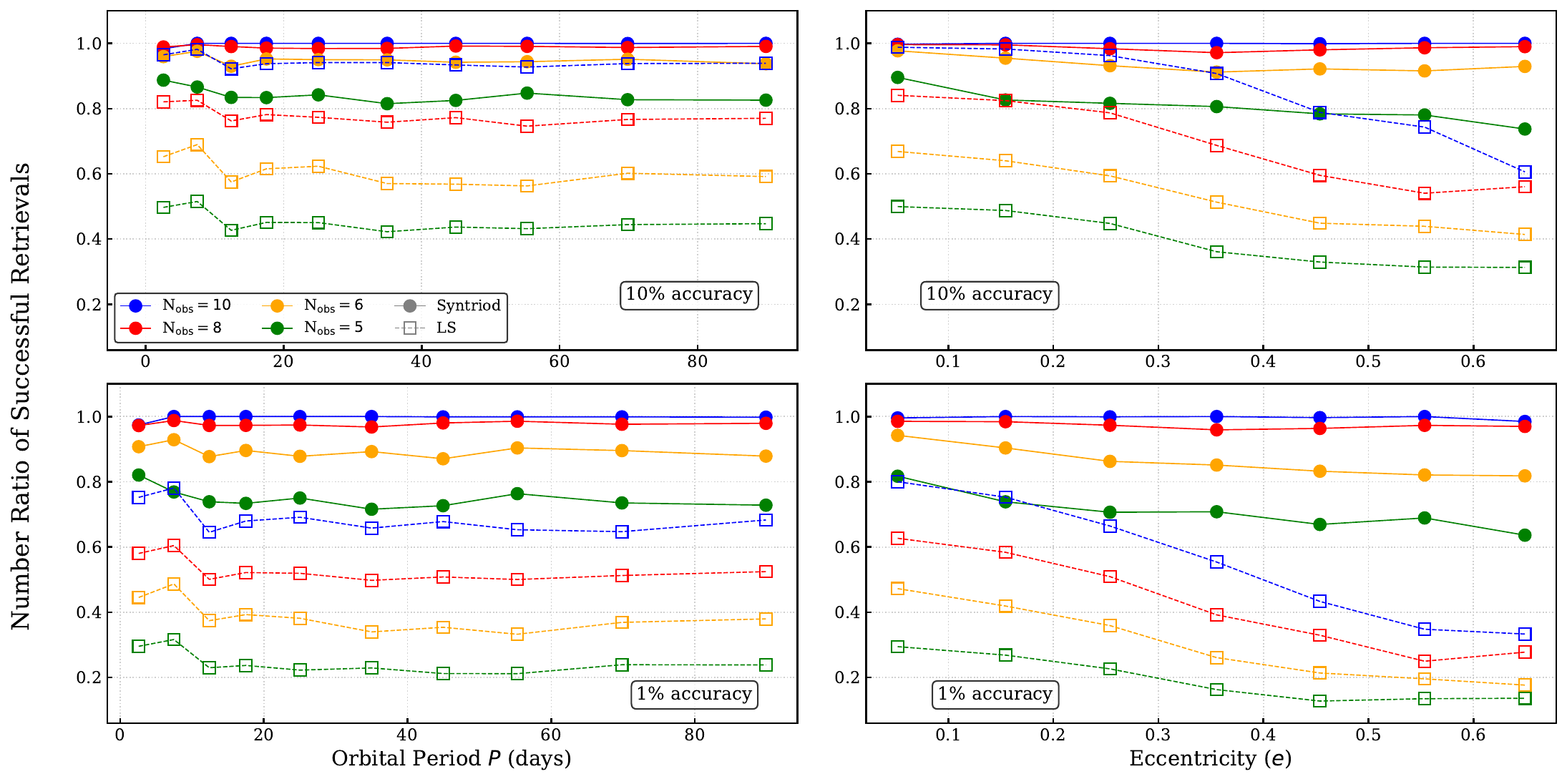}
    \caption{ Period recovery success rates of \texttt{Syntriod} (filled circles) and LS (open squares) for different $N_{\rm obs}$ values. Top and bottom panels represent the 10\% and 1\% accuracy thresholds, respectively. Performance is shown as a function of orbital period ($P$, left columns) and eccentricity ($e$, right columns), where markers indicate the mean values of the given bin.}
    \label{fig:P_e}
\end{figure*}

{Figure~\ref{fig:P_e} presents the recovery fractions obtained from 10,000 synthetic systems as functions of the true orbital period ($P$) and eccentricity ($e$) for binned data. The upper panels correspond to a recovery tolerance of 10\%, while the lower panels show the more precise 1\% criterion.
The period recovery performance of both \texttt{Syntriod} and LS remains nearly constant across the entire period range, indicating no significant dependence. However, the right panels reveal a significant difference in their response to orbital eccentricity. Since the classical LS periodogram searches only for sinusoidal signals, its performance decreases rapidly beyond $e \approx 0.3$, where the radial-velocity curves become increasingly asymmetric. In contrast, the extensive template library of \texttt{Syntriod} maintains high recovery rates even for highly eccentric systems.}

\begin{figure*}
    \centering
    \includegraphics[width=1\linewidth]{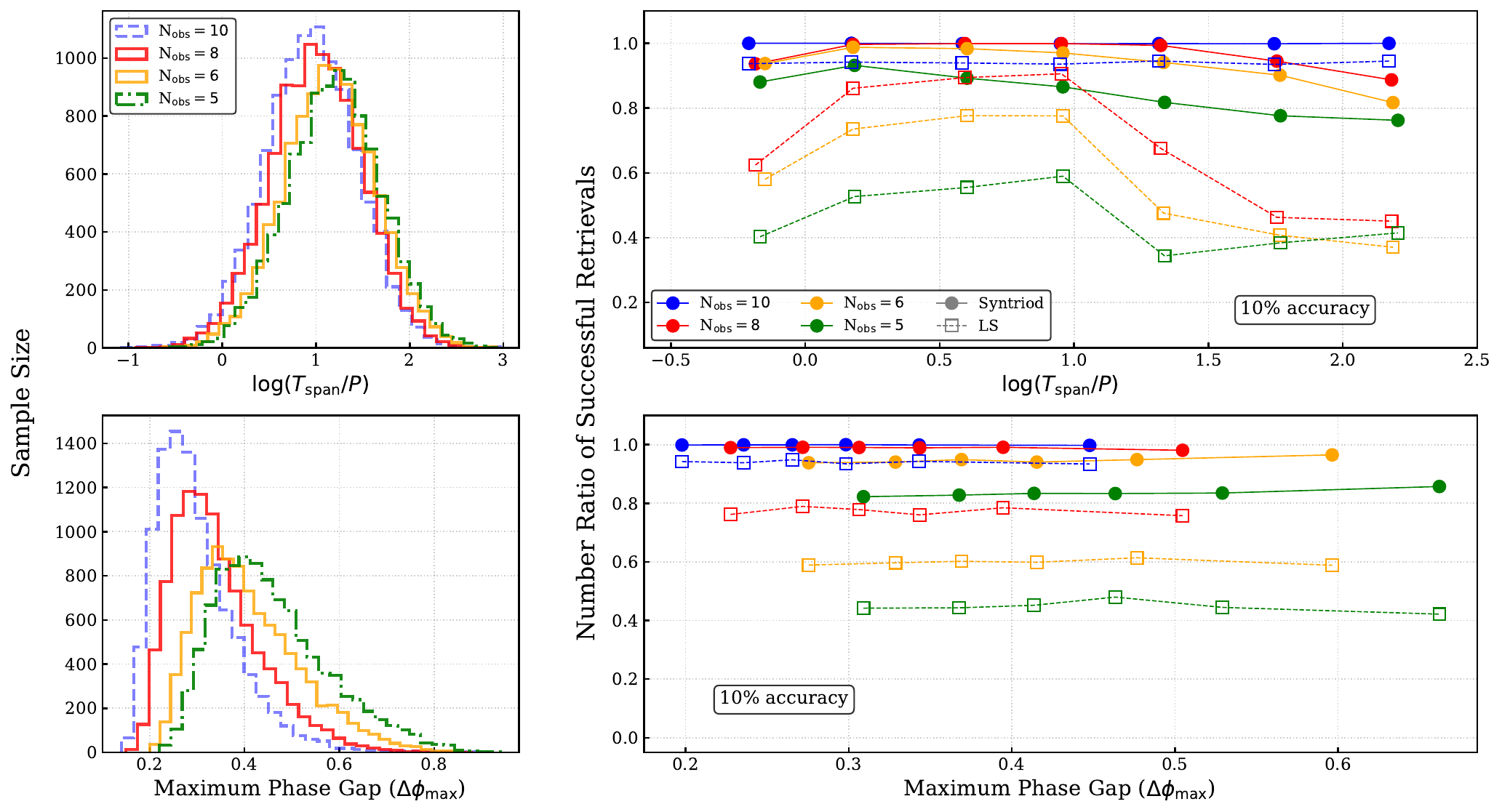}
    \caption{Observational limits and period recovery performance of \texttt{Syntriod} (filled circles) and LS (open squares) for different $N_{\rm obs}$ values. Left panels show the sample distributions of the log baseline-to-period ratio ($\log(T_{\rm span}/P)$, top) and maximum phase gap ($\Delta \phi_{\max}$, bottom). The right panels show the corresponding period recovery success rates within a 10\% accuracy threshold as a function of these parameters in a binned subsamples, where markers indicate the mean values of the given bin.}
    \label{fig:T_span_Phi}
\end{figure*}

{Figure~\ref{fig:T_span_Phi} illustrates the response of the two algorithms to two major observational challenges: long observational baselines (relative to true period) and large phase gaps, which both, as expected, slightly increase with decreasing $N_{\rm obs}$. }

{The dependence on the observational baseline reveals another important difference between the two methods. For $\log(T_{\mathrm{span}}/P)<0$, corresponding to partial-orbit regimes where the observations cover less than one orbital cycle, the recovery rate of the LS algorithm decreases noticeably, even for moderately sampled datasets ($N_{\rm obs}=8$). The LS performance also deteriorates at long baselines, with a rapid decline appearing for $\log(T_{\mathrm{span}}/P)\gtrsim1$. In contrast, the recovery fractions of \texttt{Syntriod} remain nearly uniform over the same range of observational baselines, showing only a mild downward trend toward the largest $T_{\mathrm{span}}/P$ values explored in our simulations.}

{This behavior follows naturally from the different operating domains of the two algorithms. In time-domain periodograms such as LS, a small period error accumulates into a significant phase drift when $T_{\mathrm{span}}/P$ becomes large, progressively reducing the coherence of the folded signal. The associated periodogram peaks become broader or displaced, increasing the probability of alias solutions. By contrast, \texttt{Syntriod} evaluates candidate solutions directly in the orbital-phase domain. Incorrect periods therefore produce misaligned phase-folded patterns and receive substantially lower likelihoods, largely suppressing the effect of accumulated phase drift even for long observational baselines.}

{The analysis based on the maximum phase gap ($\Delta \phi_{\max}$) reveals a more stable behavior for both methods.
Overall, these results show that \texttt{Syntriod} maintains nearly uniform recovery fractions over a broad range of orbital and sampling conditions. At the same time, the LS algorithm becomes increasingly sensitive to eccentricity and observational baseline.}

\subsection{Performance of Orbital Parameter Recovery on Synthetic Data}

Because the orbital period is the key parameter governing Keplerian solutions, the primary performance comparison in this work focuses on period recovery, particularly against the widely used Lomb–Scargle (LS) method. However, LS is designed solely for period detection, whereas \texttt{Syntriod} simultaneously estimates the full set of Keplerian orbital parameters. To evaluate the robustness of these predictions, we therefore examine the joint posterior distributions of the recovered orbital parameters.

\begin{figure*}
    \centering
    \includegraphics[width=1\linewidth]{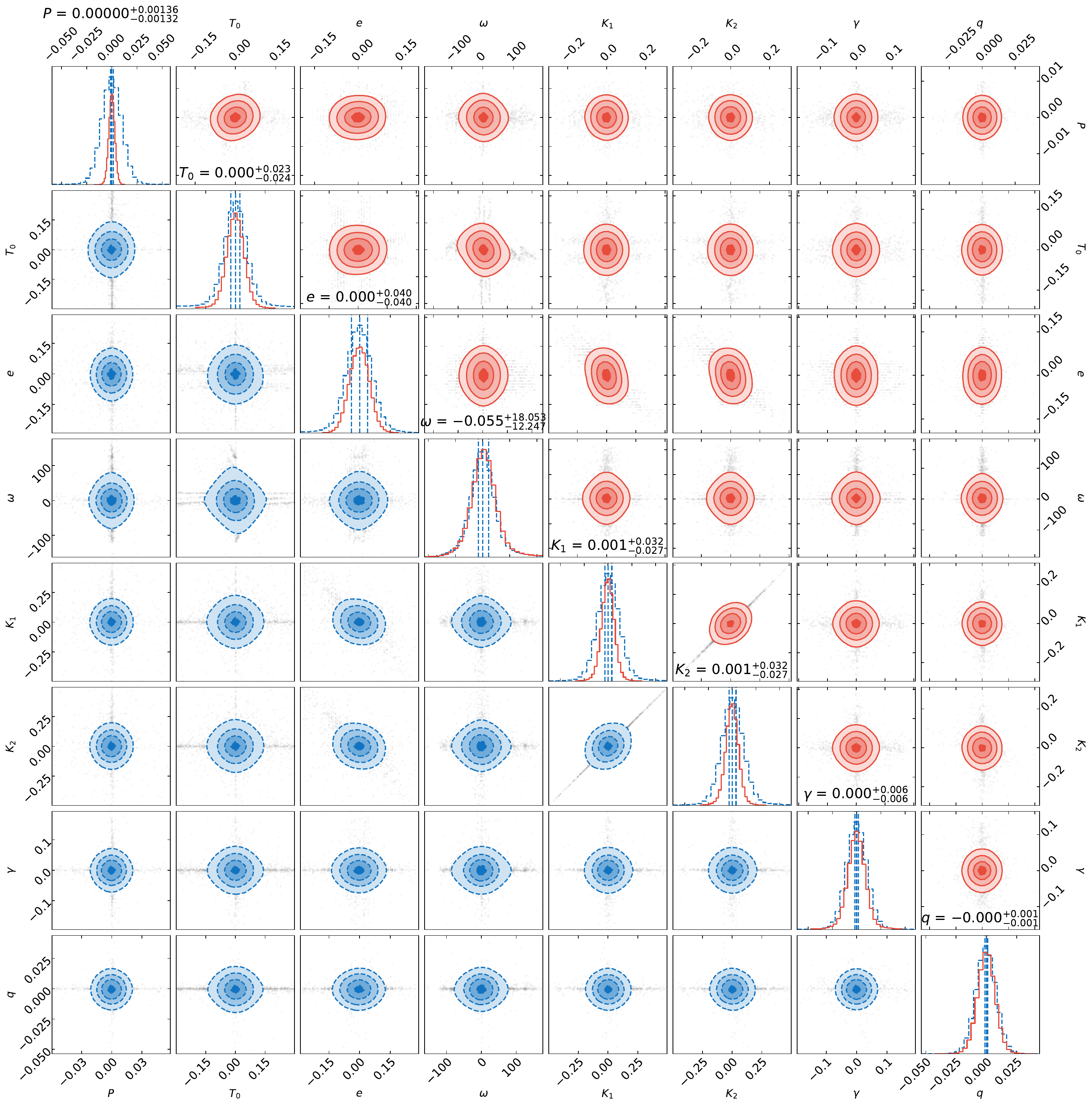}
    \caption{Corner plot showing the posterior probability distributions and covariances of the orbital parameters ($P$, $T_0$, $e$, $\omega$, $K_1$, $K_2$, $\gamma$ and $q$). The blue dashed contours and histograms represent the results obtained using 6 observational data points ($N_{obs}=6$). In contrast, the red solid contours indicate the solutions constrained by 10 data points ($N_{obs}=10$). The diagonal panels show the 1D marginalized posteriors, while the off-diagonal panels illustrate the 2D joint distributions.}
    \label{fig:corner}
\end{figure*}

{Figure~\ref{fig:corner} presents the joint posterior distributions of the Keplerian parameters ($P, T_0, e, \omega, K_1, K_2, \gamma, q$) obtained with \texttt{Syntriod} for two extreme sampling regimes: the optimal case with $N_{\rm obs}=10$ and the theoretical lower bound with $N_{\rm obs}=6$. We aim to visualize the structural impact of data sparsity on parameter recovery by combining the two scenarios in a single corner plot.}

For $N_{\rm obs}=10$, the diagonal panels show nearly Gaussian posterior distributions centered on the true values with narrow $1\sigma$ confidence intervals. The off-diagonal contours remain compact and predominantly elliptical, indicating weak parameter degeneracies and a well-constrained solution space. Physically coupled parameters, such as $K_1$ and $K_2$, preserve their expected linear correlation, while no artificial covariance structures emerge. This behavior confirms that, under sufficient phase coverage, \texttt{Syntriod} converges to a statistically stable and physically consistent orbital solution.

{In contrast, the $N_{\rm obs}=6$ case exhibits the expected broadening of posterior distributions due to the reduced number of constraints. The diagonal histograms remain centered on the true values, indicating the absence of systematic bias, although the $1\sigma$ intervals expand noticeably. The off-diagonal panels reveal increased covariance, particularly for parameters intrinsically linked through orbital geometry (e.g., $P$-$e$). Despite this widening, the contours retain coherent unimodal structures rather than fragmenting into multimodal or alias-driven solutions. This finding demonstrates that even below the theoretical sampling limit, \texttt{Syntriod} maintains a physically constrained parameter space and avoids catastrophic degeneracies.}

{The direct comparison between the $N_{\rm obs}=10$ and $N_{\rm obs}=6$ regimes therefore reveals a smooth degradation in precision rather than a structural breakdown in parameter recovery. While uncertainties naturally increase at the lower sampling limit, the overall topology of the posterior distributions remains stable, confirming that \texttt{Syntriod} preserves solution consistency even under extreme data sparsity.} {Additionally, the fact that the $\gamma$ and $q$ parameters exhibit identical distributions in both the $N_{\rm obs}=10$ and $6$ cases confirms that their precision in \texttt{Syntriod} is determined by the respective radial velocity errors rather than the sample size.}

\section{Validation and Tests on Real Spectroscopic Binary Systems}
\label{sec:real_sb}

\texttt{Syntriod}'s predictions on synthetic data (see Section~\ref{period estimation}) demonstrate its statistical performance under controlled conditions. However, synthetic datasets do not fully reproduce the complexity of real observations, which typically include measurement uncertainties, instrumental systematics, and irregular temporal sampling.
To verify that the behavior observed in the synthetic tests remains valid under realistic observing conditions, we therefore apply the same analysis framework to a set of well-characterized spectroscopic binary systems drawn from the literature. These benchmark systems span a wide range of orbital periods, eccentricities, and observational sampling patterns, allowing the algorithm's robustness to be evaluated across a broad range of parameter space.

For each system, we use the original RV measurements reported in the literature to reproduce the reference orbital solutions. In a second step, we randomly subsample the datasets to assess {potential} effects of sparse sampling, then analyze them using both \texttt{Syntriod} and the LS method, enabling a direct comparison between the two approaches under realistic observational noise and sparse-sampling conditions.

\subsection{Validation on Long-Period Spectroscopic Binaries}

The first group consists of systems with orbital periods $P > 5$ days. For these systems, the period search range is wide ($P \in [0.1,1000]$ days), consistent with the synthetic experiments described in Section~3.2. This configuration allows the algorithms to operate without a need for strong prior constraints on the orbital period. Table~\ref{fig:table1} summarizes the results for the selected long-period benchmark systems. Comparing the reference orbital parameters with the recovered solutions, \texttt{Syntriod} consistently reproduces the literature values not only for the complete datasets but also for the {independently drawn non-nested} observations with $N_{\rm obs}=10$, $6$, and $5$. {Furthermore, no significant variations were observed in the outcomes when this subsampling procedure was repeated five times.}


For the extreme example of HD~160934 \citep{griffin_ak,griffin_}, an SB1 system with $P \simeq 3748$ days and $e=0.65$, we expand our search range to be $P \in [100,5000]$ days.
\texttt{Syntriod} preserves the correct long-period prediction even when the data are reduced to $N_{\mathrm{obs}}=6$ and 5, while LS shifts to substantially different periods. A similar trend appears for Phi~Cygni \citep{Rach_cyg}, where \texttt{Syntriod} keeps the recovered period close to the reference value of $\sim434$ d across all subsampled cases, whereas LS converges to shorter alias periods. The same pattern also emerges in the eccentric Kepler binary KIC~3858884 \citep{Maceroni_kic38}: \texttt{Syntriod} remains close to the true $\sim26$ d solution in most sampling cases, while LS moves to 73.8, 36.8, and 40.2 d for $N_{\mathrm{obs}}=10$, 6, and 5, respectively. These systems show that \texttt{Syntriod} remains stable even when eccentricity and sparse phase coverage strongly deform the radial-velocity curve away from a sinusoidal shape.

In KIC~6867766 \citep{Fleming_kic68}, KIC~3003991\citep{Mahadevan_kic}, and the 5-point system KIC~2445134 \citep{Mahadevan_kic}, \texttt{Syntriod} recovers periods consistent with the real values, while LS oftentimes locks onto harmonics or long-period aliases. KIC~2445134 is particularly important because its number of observations is below the minimum-data limit (only 5), yet \texttt{Syntriod} still returns $P=8.409$ d, compared to the literature value of 8.412 d, whereas LS yields 45.36 d. Kepler-16 \citep{Bender_kepler} provides a similar example in a naturally sparse dataset, where \texttt{Syntriod} remains close to the reference solution for both $N_{\mathrm{obs}}=6$ and 5.

Across the full set of long-period benchmark systems, \texttt{Syntriod} therefore reproduces the reference orbital parameters with high consistency. At the same time, the LS solutions become increasingly sensitive to aliasing effects as the number of observations decreases.

\begin{table*}
\centering 
\setlength{\tabcolsep}{4pt}
\renewcommand{\arraystretch}{0.8}
\caption{Orbital parameter estimates obtained with \texttt{Syntriod} for long-period spectroscopic binaries, compared with reference solutions from the literature. Results are shown for the full dataset and for randomly subsampled datasets ($N_{\mathrm{obs}}= 10, 6, 5$). The final columns list total observation baselines ($T_{\mathrm{span}}$), maximum phase gaps ($\Delta \phi_{\max}$), and Lomb--Scargle period estimates.}
\begin{tabular}{@{}clc c c c c c c c c c | c@{}}
\toprule
{Binary} & {Reference}& {$N_{obs}$} & {P} & {T$_0$} & {e} & {$\omega$} & {K$_1$} & {K$_2$} & {$\gamma$} & {$T_{\mathrm{span}}$} & {$\Delta \phi_{\max}$} & {P$_{LS}$} \\ 
{System}&&&(days)& (MJD)& & ($^\circ$) & ($\mathrm{km\,s^{-1}}$) & ($\mathrm{km\,s^{-1}}$) &($\mathrm{km\,s^{-1}}$) & (days) & & (days)\\
\midrule

\multirow{6}{*}{\shortstack{{HD}\\{160934} \\ (SB1)}}
&{G13} & 82 & 3748.0 & 52429.0 & 0.65 & 218 & 7.90 & - & $-29.51$ & 6565.7 & 0.09 & - \\
&\multicolumn{11}{c|}{
\ \ \ \ \ \ \ \ \ \ \ \ \ \ \ \ \ \  - - - - - - - - - - - 
- - - - - - - - - - - - - - - - - - - - 
- - - - - - - - - - - - - - - - - - - - - - - - - -}\\
& \multirow{4}{*} {{Syntriod}}& 82 & 3769.8838 & 52419.28 & 0.70 & 220 & 8.07 & - & $-29.59$ & 6565.7 & 0.09 & 4339.1226 \\
&& 10 & 3708.8695 & 52430.81 & 0.60 & 210 & 7.79 & - & $-28.97$ & 3976.0 & 0.68 & 1811.8776 \\
&& 6 & 3589.9315 & 55907.54 & 0.70 & 150 & 9.70 & - & $-26.62$& 3424.7 & 0.69 & 2032.5599\\
&& 5 & 3529.0380 & 53193.86 & 0.50 & 20 & 14.81 & - & $-36.87$ & 4291.1 & 0.62 & 1274.1797 \\
\midrule
\multirow{6}{*} { \shortstack{{Phi}\\{Cygni}}}&{RH61} & 27 & 434.086 & 30837.64 & 0.51 & 216 & 26.79 & 27.88 & 5.00 & 3050.6 & 0.35 & - \\
&\multicolumn{11}{c|}{
\ \ \ \ \ \ \ \ \ \ \ \ \ \ \ \ \ \  - - - - - - - - - - - 
- - - - - - - - - - - - - - - - - - - - 
- - - - - - - - - - - - - - - - - - - - - - - - - -}\\
&\multirow{4}{*} {{Syntriod}}& 27 & 434.4048 & 30848.41 & 0.50 & 210 & 26.52 & 27.66 & 5.04 & 3050.6 & 0.35 & 302.0479 \\
&& 10 & 432.5395 & 30847.39 & 0.50 & 220 & 26.42 & 26.91 & 5.16 & 3030.7 & 0.49 & 306.1393 \\
&& 6 & 434.2078 & 30838.37 & 0.50 & 220 & 26.37 & 27.66 & 5.42 & 2910.0 & 0.56 & 326.9708\\
&& 5 & 424.3967 & 30889.47 & 0.50 & 210 & 28.17 & 28.83 & 5.40 & 3011.7 & 0.50 & 244.8570 \\
\midrule
\multirow{6}{*} { \shortstack{{Capella}\\ {A}}}&
{T15} &$>$ 500 & 104.0213 & 48147.60 & 0.00 & 342 & 25.90 & 26.90 & 29.90 & - & - & - \\
&\multicolumn{11}{c|}{
\ \ \ \ \ \ \ \ \ \ \ \ \ \ \ \ \ \  - - - - - - - - - - - 
- - - - - - - - - - - - - - - - - - - - 
- - - - - - - - - - - - - - - - - - - - - - - - - -}\\
&\multirow{4}{*} {{Syntriod}}& 14 & 104.1357 & 56598.88 & 0.00 & 270 & 25.95 & 26.20 & 29.70 & 446.8 & 0.22 & 103.9146 \\
&& 10 & 105.2153 & 56642.41 & 0.00 & 270 & 25.57 & 25.86 & 29.59 & 421.7 & 0.32 & 102.8732 \\
&& 6 & 102.8925 & 56905.71 & 0.00 & 270 & 25.54 & 25.66 & 29.24 & 90.7 & 0.25 & 100.8547 \\
&& 5 & 103.7717 & 56942.55 & 0.10 & 160 & 28.07 & 29.53 & 30.12 & 446.8 & 0.41 & 135.4039 \\
\midrule
\multirow{4}{*} {\shortstack{{Kepler}\\ {16}}}&{B12}& 6& 41.0778 & 57573.10 & 0.16 & 263 & 13.67 & 46.88 & $-33.00$ & 26.9 & 0.34& - \\
&\multicolumn{11}{c|}{
\ \ \ \ \ \ \ \ \ \ \ \ \ \ \ \ \ \  - - - - - - - - - - - 
- - - - - - - - - - - - - - - - - - - - 
- - - - - - - - - - - - - - - - - - - - - - - - - -}\\
&\multirow{2}{*} {{Syntriod}}
 & 6 & 39.9182 & 55847.40 & 0.10 & 260 & 13.84 & 46.35 & $-33.43$ & 26.9 & 0.34 & 38.4438 \\
&& 5 & 40.0110 & 55847.43 & 0.10 & 260 & 13.99 & 46.86 & $-33.59$ & 22.9 & 0.45 & 32.7483 \\
\midrule
\multirow{6}{*} {\shortstack{{KIC}\\ {3858884}}} &{M14}& 83 & 25.9520 & 55013.83 & 0.50 & 21 & 61.20 & 61.90 & 16.14 & 698.7 & 0.07 & - \\
&\multicolumn{11}{c|}{
\ \ \ \ \ \ \ \ \ \ \ \ \ \ \ \ \ \  - - - - - - - - - - - 
- - - - - - - - - - - - - - - - - - - - 
- - - - - - - - - - - - - - - - - - - - - - - - - -}\\
&\multirow{4}{*} {{Syntriod}}& 83 & 25.9511 & 55505.02 & 0.50 & 20 & 64.94 & 65.09 & 16.47 & 698.7 & 0.07 & 25.9746 \\
&& 10 & 25.9398 & 55556.81 & 0.40 & 20 & 57.03 & 60.00 & 15.89 & 597.4 & 0.40 & 73.7604 \\
&& 6 & 23.6120 & 55555.24 & 0.60 & 20 & 81.39 & 83.57 & 16.13 & 526.6 & 0.51 & 36.8288 \\
&& 5 & 25.9717 & 55739.20 & 0.40 & 210 & 55.30 & 57.29 & 17.55 & 366.1 & 0.65 & 40.2378 \\
\midrule
\multirow{6}{*} {\shortstack{{KIC}\\ {6867766}}}&{F15}& 20 & 12.9647 & 56746.58 & 0.05 & 128 & 30.77 & 79.40 & 10.87 & 261.0 & 0.15 & - \\
&\multicolumn{11}{c|}{
\ \ \ \ \ \ \ \ \ \ \ \ \ \ \ \ \ \  - - - - - - - - - - - 
- - - - - - - - - - - - - - - - - - - - 
- - - - - - - - - - - - - - - - - - - - - - - - - -}\\
&\multirow{4}{*} {{Syntriod}}& 20 & 12.9482 & 56567.12 & 0.00 & 0 & 30.52 & 78.67 & 10.53 & 261.0 & 0.15 & 12.9867 \\
&&10 & 12.9763 & 56563.34 & 0.00 & 270 & 30.39 & 77.80 & 10.38 & 259.0 & 0.37 & 13.0171 \\
&&6& 12.9490 & 56596.28 & 0.00 & 270 & 31.37 & 80.48 & 11.02 & 236.1 & 0.41 & 21.2724 \\
&&5 & 12.9138 & 56563.46 & 0.10 & 230 & 34.13 & 88.66 & 10.57 & 258.0 & 0.38 & 52.5989 \\
\midrule
\multirow{2}{*} {\shortstack{{KIC}\\ {2445134}}}&{M19}& 5 & 8.4120 & 55826.85 & 0.01 & 275 &37.20 & 90.00 & 21.60 & 40.9 & 0.42 & - \\
&{Syntriod}& 5 & 8.4090 & 55811.52 & 0.00 & 350 & 37.30 & 89.50& 21.02 & 40.2 & 0.42 & 45.3578 \\
\midrule
\multirow{6}{*} {\shortstack{{KIC}\\ {3003991}}}&{M19}& 12 & 7.2448 & 55953.65 & 0.00 & 234 & 24.97 & 83.00 & $-122.60$ & 275.0 & 0.23 & - \\
&\multicolumn{11}{c|}{
\ \ \ \ \ \ \ \ \ \ \ \ \ \ \ \ \ \  - - - - - - - - - - - 
- - - - - - - - - - - - - - - - - - - - 
- - - - - - - - - - - - - - - - - - - - - - - - - -}\\
&\multirow{4}{*} {{Syntriod}}&12 & 7.2453 & 55817.64 & 0.10 & 170 & 25.99 & 88.47 & $-121.71$ &275.0 & 0.23 & 14.5539 \\
&& 10& 7.2447 & 55819.73 & 0.00 & 270 & 25.27 & 85.77 & $-121.60$ & 241.3 & 0.26 & 14.5539 \\
&& 6 & 7.2439 & 55814.28 & 0.00 & 0 & 24.84 & 88.26 & $-122.46$ & 273.1 & 0.30 & 36.4133 \\
&& 5 & 7.2857 & 55842.99 & 0.10 & 160 & 24.29 & 81.67 & $-122.19$ & 244.1 & 0.40 & 28.0585 \\
\bottomrule
\end{tabular}

\vspace{1ex} 
\raggedright
\label{fig:table1}
\textit{Notes for references: G13; \cite{griffin_}, RH61; \cite{Rach_cyg}, T15; \cite{Torres_capella}, B12; \cite{Bender_kepler}, M14; \cite{Maceroni_kic38}, F15; \cite{Fleming_kic68}, M19; \cite{Mahadevan_kic}} 
\end{table*}

\subsubsection{Validation on Short-Period Spectroscopic Binaries}

The second group consists of short-period systems with $P \leq 1$ day. These binaries often exhibit stronger aliasing and harmonic ambiguities due to their rapid orbital motion, making them particularly challenging for general period-search algorithms.

To explore this regime, the period search was performed under two configurations: a broad search interval ($P \in [0.1,100]$ days) and a narrower interval ($P \in [0.01,10]$ days), incorporating limited prior information about the expected period range. The results obtained under these configurations are summarized in Table~\ref{fig:table2}.

The analysis includes several representative systems with diverse physical characteristics, including DU~Boo \citep{Pribulla}, HL~Dra \citep{Pribulla}, FP~Boo \citep{Rucinski}, and AK~Her \citep{Pribulla}. Despite the complexity of these systems, \texttt{Syntriod} successfully recovers the correct orbital period in the majority of tested scenarios, even with reduced datasets of $N_{\rm obs}=5$ or $6$.
For example, in the case of DU~Boo ($P\approx1.055$ days), \texttt{Syntriod} consistently identifies the correct period across all sampling levels, whereas the LS method frequently converges to alias solutions, particularly for sparse datasets. Similarly, the SB1 system HL~Dra is successfully modeled by \texttt{Syntriod} using only the primary velocity component, while LS solutions tend to lock onto incorrect harmonic periods when for subsampled  RV observations. AK~Her is especially notable: \texttt{Syntriod} keeps the solution within 0.4211--0.4234 d for all tested subsets, while LS collapses to clear aliases at 0.0119, 0.1401, and 0.1166 d.
Even for challenging systems such as FP~Boo, which has an extremely low mass ratio ($q\approx0.1$), \texttt{Syntriod} can still yield orbital parameters close to the literature values once $N_{\rm obs}\ge6$. The contact binary AK~Her likewise demonstrates stable parameter recovery across all tested sampling levels.

\begin{table*}
\centering 
\setlength{\tabcolsep}{4pt}
\renewcommand{\arraystretch}{0.8}
\caption{Orbital parameter estimates obtained with \texttt{Syntriod} for short-period ($P \le 1$ day) spectroscopic binaries, compared with reference solutions from the literature. Results are shown for the full dataset and for randomly subsampled datasets ($N_{\mathrm{obs}}= 10, 6, 5$). The final columns list total observation baselines ($T_{\mathrm{span}}$), maximum phase gaps ($\Delta \phi_{\max}$), and Lomb--Scargle period estimates.}
\begin{tabular}{@{}clc c c c c c c c c c | c@{}}
\toprule
{Binary} & {Reference}& {$N_{obs}$} & {P} & {T$_0$} & {e} & {$\omega$} & {K$_1$} &{K$_2$} & {$\gamma$} & {$T_{\mathrm{span}}$} & {$\Delta \phi_{\max}$} & {P$_{LS}$} \\ 
{System} &&&(days)& (MJD)& & ($^\circ$) & ($\mathrm{km\,s^{-1}}$) & ($\mathrm{km\,s^{-1}}$) &($\mathrm{km\,s^{-1}}$) & (days) & & (days)\\
\midrule
\multirow{6}{*}{{DU Boo}} &{P6} &27  & 1.0558 &  53494.68 & 0.00 & - & 53.59 &  229.32 & $-13.09$ & 31.1 & 0.25 & - \\
&\multicolumn{11}{c|}{
\ \ \ \ \ \ \ \ \ \ \ \ \ \ \ \ \ \  - - - - - - - - - - - 
- - - - - - - - - - - - - - - - - - - - 
- - - - - - - - - - - - - - - - - - - - - - - -}\\
 &\multirow{4}{*}{{Syntriod}}& 27 & 1.0558 & 53481.52 & 0.10 & 270 &  54.25 & 232.65 &  $-13.36$ & 31.1 & 0.25 & 1.9833 \\
 & & 10 & 1.0519 & 53481.52& 0.10 & 270 & 53.53 &  230.74 & $-11.24$ & 31.1 & 0.32 & 0.1131 \\
 && 6 &1.1434 & 53481.58 & 0.10 & 270 & 53.49 & 226.13 & $-11.53$ & 27.1 & 0.36 & 0.1003 \\
 && 5 & 1.0556 & 53480.89 &  0.10 & 240 & 55.80 & 227.43 & $-16.70$ & 31.0 & 0.54 & 0.1615 \\
\midrule
\multirow{6}{*}{\shortstack{{HL Dra} \\ (SB1)}} &{P6} &126  & 0.9443 & 53166.81 & 0.00 & - & 81.04 & - & $-37.00$ & 339.1 & 0.04 & - \\
&\multicolumn{11}{c|}{
\ \ \ \ \ \ \ \ \ \ \ \ \ \ \ \ \ \  - - - - - - - - - - - 
- - - - - - - - - - - - - - - - - - - - 
- - - - - - - - - - - - - - - - - - - - - - - -}\\
&\multirow{4}{*}{{Syntriod}}& 126 & 0.9443 & 53162.55 & 0.00 & 270 & 80.67 & - & $-34.04$ & 339.1 & 0.04 & 0.9470 \\
&& 10 &0.9386 & 53168.96 & 0.10 & 10 & 77.54 & - & $-37.86$ & 313.1 & 0.24 & 0.9354 \\
&&6 &{*}0.9446 & 53460.84 & 0.00 & 270 & 79.54& - & $-31.82$ & 21.9 & 0.41& 1.7997 \\
&& 5 & 0.9514 &  53169.50 & 0.00 & 0 & 87.77 & - &  $-42.43$ & 26.7 & 0.71 & 0.1341 \\
\midrule
\multirow{6}{*}{{FP Boo}}&{R5} & $\sim$25  & 0.6405 & 52388.22 & 0.00 & - & 26.80 & 254.04 & $-4.87$ & 421.8 & 0.29 & - \\
&\multicolumn{11}{c|}{
\ \ \ \ \ \ \ \ \ \ \ \ \ \ \ \ \ \  - - - - - - - - - - - 
- - - - - - - - - - - - - - - - - - - - 
- - - - - - - - - - - - - - - - - - - - - - - -}\\
&\multirow{4}{*}{{Syntriod}}& 20 & {*}0.6395 &52385.75 & 0.10 & 150 & 26.57 & 263.89 & $-10.58$ & 421.8 & 0.29 & 0.5203  \\
&& 10 & 0.6415 & 52385.96 & 0.10 & 270 & 25.60 &  270.81 & $-8.32$ & 421.8 & 0.35 & 0.3806\\
&& 6 & 0.6395& 52386.33 & 0.10 & 110 &  25.32 & 250.32 & $-4.03$ & 396.8 & 0.45 & 0.8008 \\
 &&5 & {*}0.6573 & 52385.91 & 0.20 & 260 & 24.06 &  249.67 & $-6.42$ & 382.8 & 0.47 & 0.1003 \\
\midrule
\multirow{6}{*}{{AK Her}}&{P6}& 37 & 0.4215 & 53176.39 & 0.00 & - & 70.52 & 254.40 & 4.28 & 93.8 & 0.27 & - \\
&\multicolumn{11}{c|}{
\ \ \ \ \ \ \ \ \ \ \ \ \ \ \ \ \ \  - - - - - - - - - - - 
- - - - - - - - - - - - - - - - - - - - 
- - - - - - - - - - - - - - - - - - - - - - - -}\\
&\multirow{4}{*}{{Syntriod}}&37 & 0.4215 & 53159.83 & 0.00 & 350 & 70.31&  260.17 &   6.16 & 93.8 & 0.27 & 0.4215  \\
&& 10 & 0.4211 & 53160.17 & 0.10 & 270 & 65.69 & 254.94&   3.67 & 23.0 & 0.36 & 0.1119 \\
 && 6 & 0.4227 & 53159.95 & 0.10 & 100 & 72.17 & 274.59 & 6.83 & 29.9 & 0.43 & 0.1401 \\
 && 5 & 0.4234 & 53159.92 & 0.10 & 210 & 72.52 & 265.42 & 5.59 & 7.8 & 0.51 & 0.1166 \\
\bottomrule
\end{tabular}

\vspace{1ex} 
\raggedright
\textit{*} \texttt{Syntriod} achieved the optimal solution within the narrower $P \in [0.01, 10]$ days range.

\textit{Notes for references: P6; \cite{Pribulla}, R5; \cite{Rucinski}}
\label{fig:table2}
\end{table*}

Overall, these results confirm that \texttt{Syntriod} maintains reliable performance across a diverse range of short-period systems, while traditional period-search approaches become increasingly susceptible to aliasing as observational sampling becomes sparse.

\section{Estimating \texorpdfstring{$q$ and $\gamma$}{q and gamma} Beyond The Sampling Limit}

In the final stage of the analysis, we examine the regime in which the number of observations falls below the minimum degrees of freedom required for a conventional Keplerian orbital solution ($N_{\rm obs}<5$). We find that the period-recovery success rate decreases rapidly once this limit is crossed, as expected. For $N_{\rm obs}=4$, \texttt{Syntriod} achieves a {reasonable} success rate of $\sim60\%$, while the LS method drops to $\sim35\%$ with a 10\% tolerance. This finding indicates that the predictive power of both approaches deteriorates significantly, with the LS results approaching a near-random solution regime. Although \texttt{Syntriod} may still recover the correct period in some cases, the outcome becomes strongly dependent on the phase distribution of the available observations. Consequently, without strong prior constraints or favorable sampling geometry, \texttt{Syntriod} cannot be considered as a reliable full-orbit solver at $N_{\rm obs}<5$.

\begin{figure}
    \centering
    \includegraphics[width=1\linewidth]{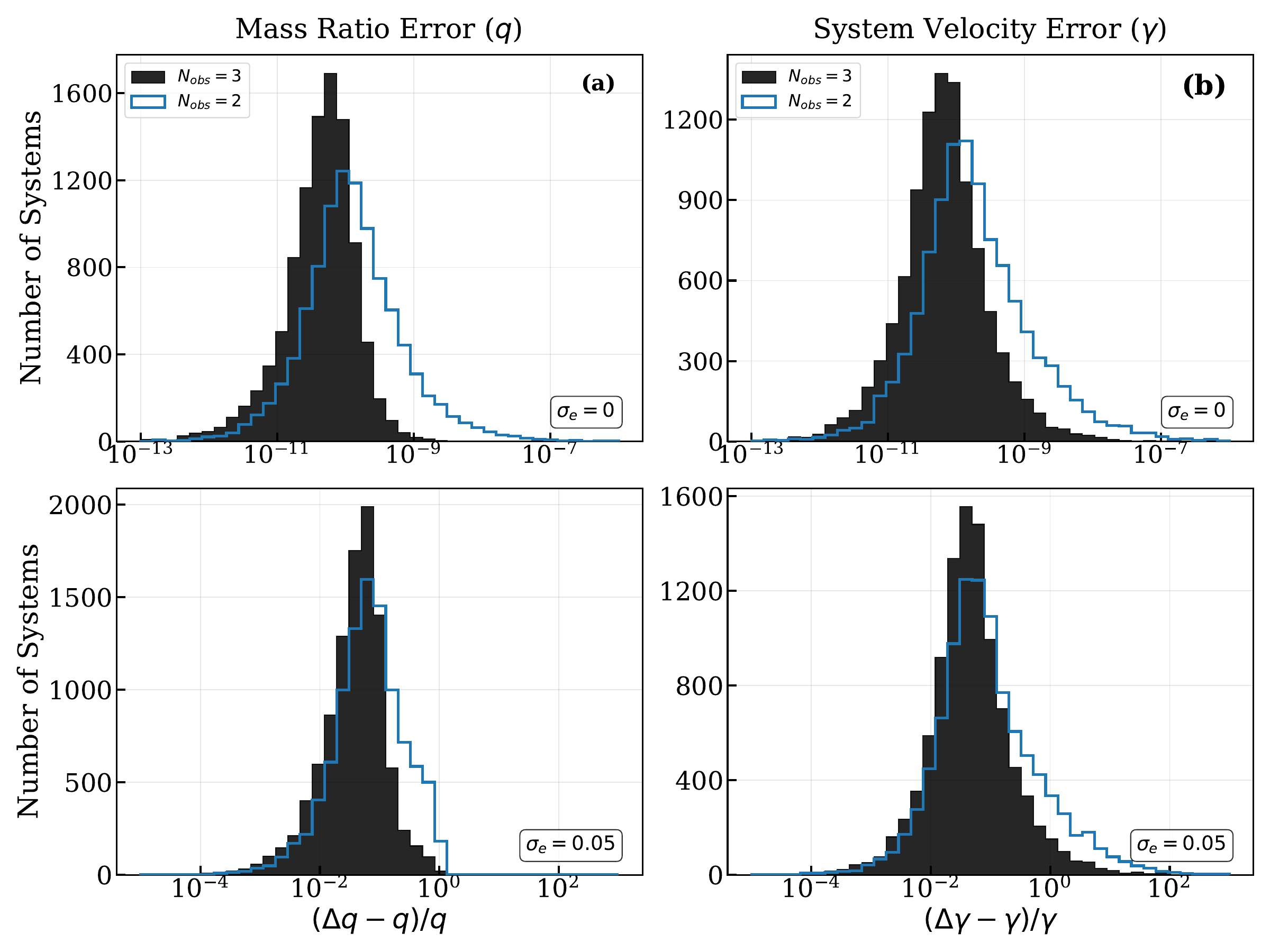}
    \caption{Relative error distributions of the mass ratio ($q$; a) and systemic velocity ($\gamma$; b) for 10\,000 simulated SB2 systems with extremely sparse observations. Each column compares ideal ($\sigma_e = 0$, top) and noisy ($\sigma_e = 0.05$, bottom) conditions. Black filled histograms represent $N_{obs}=3$, while blue solid lines denote $N_{obs}=2$.}
    \label{fig:qandgamma_1}
\end{figure}

{Recovery success of $q$ and $\gamma$ for 10,000 synthetic SB2 systems ($N_{\text{obs}} = 2, 3$) under ideal ($\sigma_e = 0$) and noisy (Gaussian distribution scaled to $\sigma_e = 0.05 \times \text{RV}$) conditions are in Figure~\ref{fig:qandgamma_1}. Estimated parameters cluster tightly around true values in ideal cases. For noisy data, $q$ is recovered within a 10\% acurracy threshold for 90\% ($N_{\text{obs}}=3$) and 73\% ($N_{\text{obs}}=2$) of systems, rising to 85\% within a 20\% threshold for the latter. The $\gamma$ parameter shows lower accuracy, with 72\% of orbits falling within the 20\% threshold. These results underscore the algorithm's capability to constrain orbital parameters even from highly sparse data coverage.}


\begin{figure}
    \centering
    \includegraphics[width=1\linewidth]{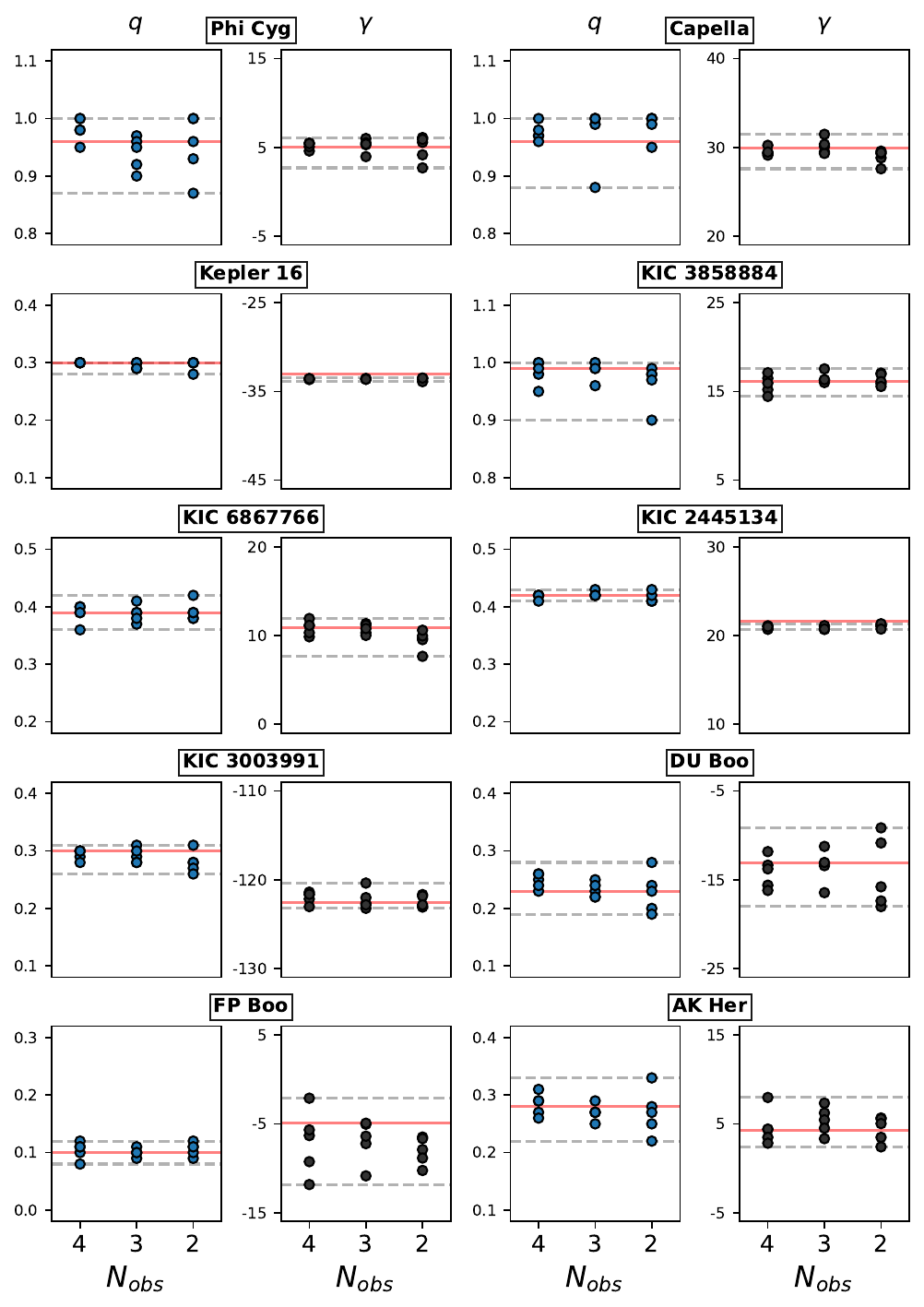}
    \caption{Verification of the mass ratio ($q$, blue) and systemic velocity ($\gamma$, black) calculations by the \texttt{Syntriod} algorithm in the extremely sparse data regime ($N_{obs} < 5$). The results correspond to the binary systems selected from the literature and analyzed in Tables 1 and 2.}
    \label{fig:qandgamma}
\end{figure}

Figure \ref{fig:qandgamma} presents the recovered mass ratio ($q$) and systemic velocity ($\gamma$) values obtained with \texttt{Syntriod} using subsets of $N_{\mathrm{obs}} = 4$, $3$, and $2$ spectroscopic observations for the SB2 benchmark systems. For each $N_{\mathrm{obs}}$, the solution was derived from randomly selected subsets of the available spectra, and the procedure was repeated five times to reduce the dependence of the results on the specific choice of observations. Even with such small numbers of observations, \texttt{Syntriod} can recover both parameters with good accuracy, generally within the uncertainties of the literature values. 

\section{Discussion}
\label{sec:discussion}

The results of this study suggest that orbital-phase domain  RV template matching offers a robust alternative for initial parameter estimation in RV curve solutions, particularly in observational regimes where sampling constraints and aliasing effects limit traditional methods.

From a practical perspective, \texttt{Syntriod}'s role differs from that of full-orbit-fitting frameworks. The method is not designed to replace Bayesian inference tools such as MCMC or nested sampling, but to complement them by providing reliable initial parameter estimates. In high-dimensional parameter spaces, these inference methods can suffer from slow convergence and sensitivity to local minima when the initial conditions are poorly constrained. By restricting the parameter space to physically plausible regions, \texttt{Syntriod} can significantly improve both computational efficiency and convergence behavior.

A further limitation of the current implementation is that \texttt{Syntriod} does not provide formal uncertainty estimates for the derived orbital parameters. While the method delivers physically consistent point estimates, it does not sample the posterior distribution and therefore cannot quantify parameter uncertainties in a statistical sense. 
This limitation is inherent to the template--matching approach and motivates using \texttt{Syntriod} in combination with full Bayesian inference methods when uncertainty quantification is required.

A key conceptual aspect of \texttt{Syntriod} is its dynamic behavior in underconstrained regimes. This structural advantage becomes particularly evident in eccentric systems. In highly non-sinusoidal radial-velocity curves, such as those observed in eccentric binaries, harmonic ambiguities can easily mislead purely Fourier-based methods. The template-based approach of \texttt{Syntriod} preserves the physical shape of the Keplerian signal and therefore remains stable even when the number of observations is close to the theoretical sampling limit. 
Surveys such as SDSS \citep{SDSS}, LAMOST \citep{Lamost}, DESI \citep{desi}, and \textit{GAIA} produce a limited number of RV  measurements for millions of stars, among which a significant fraction are spectroscopic binaries. In this context, \texttt{Syntriod} can serve as a computationally efficient pre-solver that constrains the orbital parameter space close to the physically relevant solution.

When the number of observations becomes insufficient to fully determine the Keplerian parameter space, the algorithm transitions from full orbital reconstruction to physically constrained linear relations within the radial velocity data. This approach enables robust estimates of the mass ratio ($q$) and systemic velocity ($\gamma$) even when the orbital period is poorly constrained.
Photometric missions such as \textit{Kepler} \citep{kepler} and \textit{TESS} routinely identify binary candidates with well-determined orbital periods but limited spectroscopic follow-up.  \texttt{Syntriod} provides a practical means of extracting meaningful dynamical constraints from these sparse datasets, effectively bridging the gap between photometric solution and mass estimation.  

\section{Summary and Conclusions}

In this work, we introduced \texttt{Syntriod}, an orbital-phase domain RV template-based algorithm that provides robust orbital parameter estimates from radial velocity data across a wide range of observational conditions. The method's performance was evaluated using both synthetic datasets and real spectroscopic binary systems from the literature.
The main results of this study can be summarized as follows:
\begin{itemize}
    \item \texttt{Syntriod} achieves high accuracy in period estimation across different sampling regimes. For well-sampled datasets ($N_{\mathrm{obs}} \geq 8$), the success rate exceeds $99\%$, with median relative errors at the level of $\sim 10^{-3}$.
    
    \item At the theoretical sampling limit ({${N_{\mathrm{obs}} = 6}$}), the algorithm maintains a success rate of $\sim 95\%$, significantly outperforming the Lomb--Scargle method, which degrades to $\sim 65\%$ due to aliasing.
    
    \item Below the theoretical sampling limit ($N_{\mathrm{obs}} = 5$), \texttt{Syntriod} continues to recover correct solutions in $\sim 84\%$ of cases, while Lomb--Scargle approaches the random-selection limit ($\sim 46\%$).
    
    \item Unlike time-domain approaches, \texttt{Syntriod} produces predominantly unimodal solution distributions and avoids alias-driven secondary populations. Performance degradation manifests primarily as controlled uncertainty broadening rather than catastrophic solution failure.

\item {For systems with a sufficient number of observations (particularly $N_{\rm obs}>6$), \texttt{Syntriod} maintains nearly uniform period-recovery fractions over a broad range of orbital geometries and sampling conditions, including highly eccentric orbits, large  $T_{\rm span}/P$, and large phase gaps. In contrast, the recovery performance of the LS algorithm becomes increasingly sensitive to eccentricity and observational baseline.}

\item The validation on real spectroscopic binaries confirms that this behaviour extends beyond synthetic datasets. \texttt{Syntriod} reproduces literature solutions across a broad range of systems, from very short-period contact binaries ($P \sim 0.42$ d) to long-period eccentric systems ($P \sim 3748$ d, $e \sim 0.7$), and remains consistent even when the observations are randomly subsampled to $N_{\mathrm{obs}}=5$--6.

\item In double-lined systems, \texttt{Syntriod} successfully recovers physically coupled parameters such as the mass ratio ($q$) and systemic velocity ($\gamma$), even when the number of  RV observations is limited to only 2. 
\end{itemize}

From a computational perspective, \texttt{Syntriod} remains efficient despite its exhaustive template-search strategy. On a standard computer, {for a binary orbit}, the runtime is typically ${<10}$ seconds for a few million iterations. Even on older hardware, complete solutions are obtained within $< 1$ minute (with the standard template set). 

{Although \texttt{Syntriod} demonstrates robust performance even with sparse datasets, results obtained under conditions of large RV uncertainties and limited data points ($N_{\rm obs} = 5\text{--}6$) must be interpreted with caution.}
{Despite \texttt{Syntriod}'s robustness against sparsity, outputs for observations with high RV errors and $N_{\rm obs} = 5\text{--}6$ warrant caution, particularly at high eccentricities. Additionally, aliasing effects are more prone to occur for periods near or below 1 day.}

The development of a full orbital modeling framework that builds upon the \texttt{Syntriod} parameter estimation scheme is currently ongoing and will be presented in a forthcoming study \citep[see][]{barbaros}.

\begin{acknowledgments}
{We sincerely thank the anonymous referee for their highly constructive comments and insightful suggestions, which significantly improved the quality and clarity of this manuscript.}
We thank the Scientific and Technological Research Council of Türkiye (TÜBİTAK) for financial support under project number 125F363. 

\end{acknowledgments}

\begin{contribution}

All authors contributed to the development of the \texttt{Syntriod} algorithm and the design of the study, as well as to the interpretation of the results and the writing and revision of the manuscript.



\end{contribution}



\section*{Software and Data Availability}

The \texttt{Syntriod} algorithm developed in this study is open-source and freely available on GitHub at \url{https://github.com/EmreBarbaros/Syntriod}. The synthetic datasets generated for the performance analysis and the configuration files used to reproduce the figures in this article are also available in the same repository.

This research made use of \texttt{Python} \citep{vanrossum2009} and its scientific ecosystem, including \texttt{NumPy} \citep{harris2020}, \texttt{Matplotlib} \citep{hunter2007}, \texttt{SciPy} \citep{virtanen2020},\texttt{Pandas} \citep{mckinney2010}, \texttt{Joblib} \citep{varoquaux2009}, and the JIT compiler \texttt{Numba} \citep{lam2015}.

\section*{AI Usage Statement}

The authors used AI-based language tools to assist with English editing and clarity improvement. All scientific content, analysis, and conclusions were developed and verified by the authors.

\appendix

\section{{Tests with different templates}}\label{B1}

The standard template set employed in the general application of \texttt{Syntriod} features step sizes of $\Delta e = 0.1$ and $\Delta\omega = 10^\circ$. All results presented in the Results section were computed based on this standard template. To comparatively evaluate the performance of the \texttt{Syntriod} algorithm across different template configurations, we obtained results using 1,000 representative orbits selected from a pool of 10,000 synthetic orbits. Template Set 1 ($\Delta e = 0.1, \Delta\omega = 5^\circ$) utilizes a more precise $\omega$ resolution compared to the standard set. Template Set 2 employs an adaptive step size, where $\Delta e = 0.1$ with $\Delta\omega = 10^\circ$ for $e \le 0.3$, and $\Delta\omega = 5^\circ$ for $e > 0.3$. Lastly, Template Set 3 can be used to achieve a much higher-precision estimate with $\Delta e = 0.05$ and $\Delta\omega = 5^\circ$. Since adopting smaller step sizes significantly increases the total number of templates, it consequently increases the computational runtime of \texttt{Syntriod}. 

In Table \ref{tab:b1}, we present a comparative analysis of the results from the standard template and alternative template sets across 1,000 synthetic data points for $N_{obs} = 8, 6, \text{ and } 5$. In these comparisons, the ratios of periods determined within 10\% and 1\% accuracy are reported. 

All tested \texttt{Syntriod} template libraries significantly outperform LS, particularly for sparse and eccentric systems. Template Set 3 yields the highest recovery fractions, but the differences among the template sets remain small compared with the overall advantage of \texttt{Syntriod} over LS.

\begin{table*}
\caption{Period recovery success rates of \texttt{Syntriod} using different template sets and LS for synthetic data with varying $N_{obs}$ (\%)}
\centering
\setlength{\tabcolsep}{10pt}
\renewcommand{\arraystretch}{1.2}
\begin{tabular}{c l c c | c c | c c | c c | c c}
\hline
\hline
\multirow{2}{*}{{$N_{obs}$}} & \multirow{2}{*}{{Template Set}} & \multicolumn{2}{c|}{{All Solution}} & \multicolumn{2}{c|}{{$e<0.3$}} & \multicolumn{2}{c|}{{$e>0.3$}} & \multicolumn{2}{c|}{{$T_{span}/P<10$}} & \multicolumn{2}{c}{{$T_{span}/P>10$}} \\
&& \multicolumn{2}{c|}{(1000)} & \multicolumn{2}{c|}{(765)} & \multicolumn{2}{c|}{(235)} & \multicolumn{2}{c|}{} & \multicolumn{2}{c}{} \\
\cline{3-12}
& & \%10 & \%1 & \%10 & \%1 & \%10 & \%1 & \%10 & \%1 & \%10 & \%1 \\
\hline
\multirow{5}{*}{{8}} 
& Standard & 98.1 & 96.2 & 98.7 & 96.1 & 96.2 & 93.2 & 99.4 & 96.1 & 96.7 & 96.3 \\
& Template Set 1& 97.9 & 95.8 & 98.3 & 96.2 & 96.6 & 92.8 & 99.4 & 95.5 & 96.3 & 96.1 \\
& Template Set 2 & 97.7 & 95.6 & 97.9 & 95.6 & 97.4 & 92.8 & 99.4 & 95.7 & 96.1 & 95.9 \\
& Template Set 3& 98.2 & 96.8 & 98.3 & 96.7 & 97.9 & 93.6 & 99.4 & 97.5 & 96.3 & 96.1 \\
& LS & 69.8 & 43.7 & 74.8 & 48.0 & 53.6 & 26.4 & 85.3 & 42.4 & 53.7 & 44.1 \\
\hline
\multirow{5}{*}{{6}} 
& Standard & 93.5 & 88.9 & 94.2 & 91.8 & 91.1 & 85.5 & 99.2 & 90.5 & 89.9 & 85.8 \\
& Template Set 1 & 94.0 & 91.2 & 94.9 & 92.8 & 91.1 & 89.8 & 97.7 & 91.8 & 91.7 & 90.0 \\
& Template Set 2 & 94.1 & 90.9 & 94.6 & 92.3 & 92.3 & 89.8 & 96.9 & 91.5 & 92.3 & 89.9 \\
& Template Set 3 & 94.5 & 92.4 & 95.2 & 93.5 & 93.2 & 92.3 & 98.5 & 94.6 & 92.0 & 90.4 \\
& LS & 57.1 & 35.2 & 61.8 & 39.6 & 43.0 & 20.9 & 73.5 & 39.2 & 46.4 & 32.5 \\
\hline
\multirow{5}{*}{{5}} 
& Standard & 81.5 & 73.2 & 82.1 & 73.5 & 79.1 & 69.8 & 92.9 & 79.8 & 76.5 & 69.1 \\
& Template Set 1 & 81.9 & 75.5 & 82.5 & 77.1 & 80.0 & 74.5 & 91.2 & 86.6 & 78.1 & 70.4 \\
& Template Set 2& 82.4 & 74.0 & 83.0 & 74.9 & 80.4 & 74.5 & 93.2 & 84.1 & 77.9 & 70.4 \\
& Template Set 3& 83.0 & 77.2 & 83.7 & 79.7 & 80.9 & 72.3 & 94.0 & 87.8 & 78.2 & 71.9 \\
& LS & 42.4 & 22.3 & 45.5 & 25.1 & 35.3 & 13.6 & 50.3 & 26.1 & 39.2 & 20.5 \\
\hline
\end{tabular}

\vspace{1ex}
\raggedright \footnotesize 
$^\ast$Total counts for $T_{span}/P<10$ are 510, 388, and 352 for $N_{obs}=8, 6$, and $5$ respectively. \\
$^{\ast\ast}$Total counts for $T_{span}/P>10$ are 490, 612, and 638 for $N_{obs}=8, 6$, and $5$ respectively.
\label{tab:b1}
\end{table*}

\section{{Tests on Uncertainty added synthetic data}}\label{A1}

To evaluate the robustness of \texttt{Syntriod} against measurement uncertainties, we repeated the injection--recovery experiments for 1,000 representative systems selected from the original set of 10,000 synthetic binaries after adding Gaussian noise to the radial velocities. We considered both heteroscedastic and homoscedastic uncertainties with noise levels of  $\sigma/|RV|=0.01$, $0.05$, $0.1$, and $0.2$, together with an additional fixed uncertainty of $\sigma=2$.

\begin{figure}
    \centering
    \includegraphics[width=0.9\linewidth]{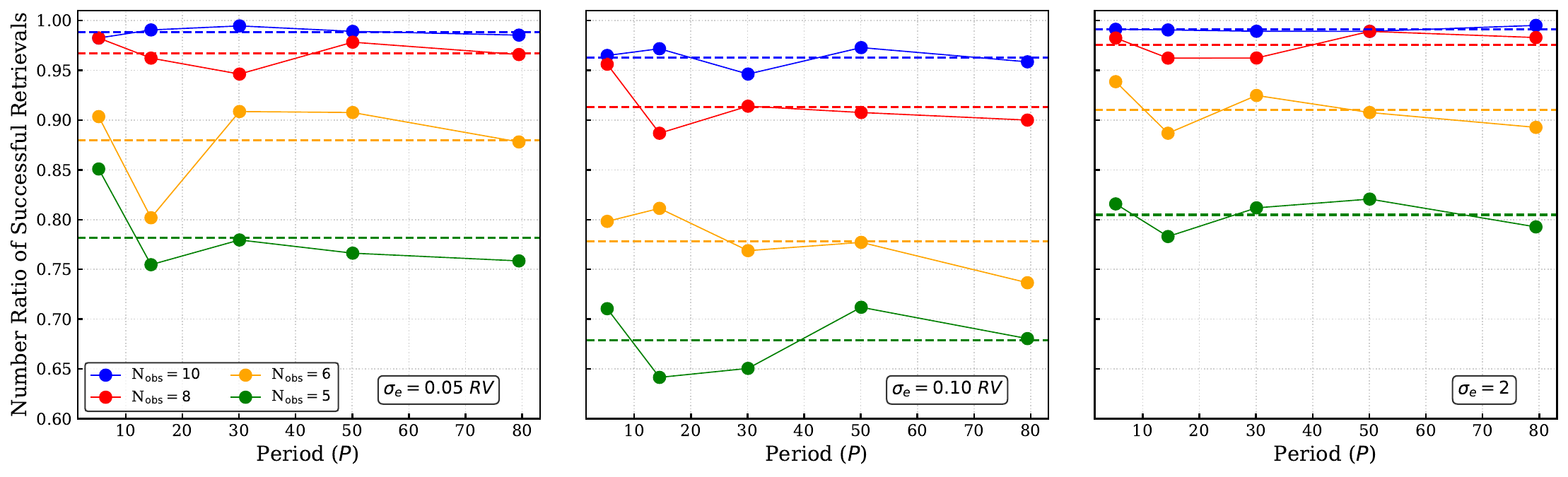}
    \includegraphics[width=0.9\linewidth]{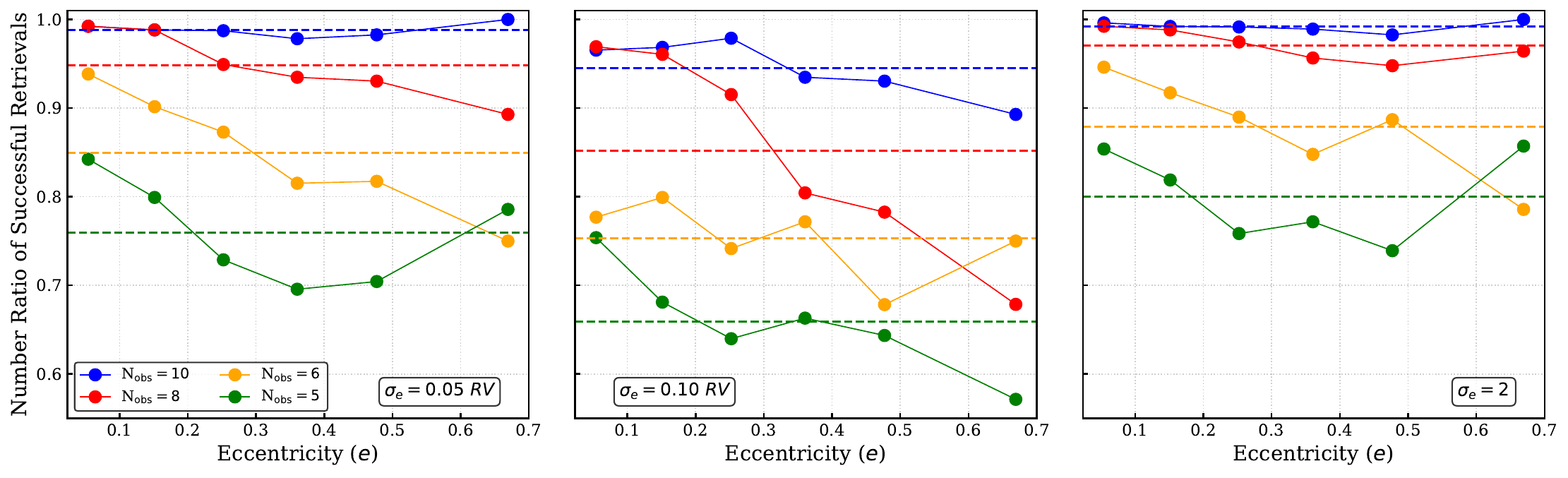}
    \includegraphics[width=0.9\linewidth]{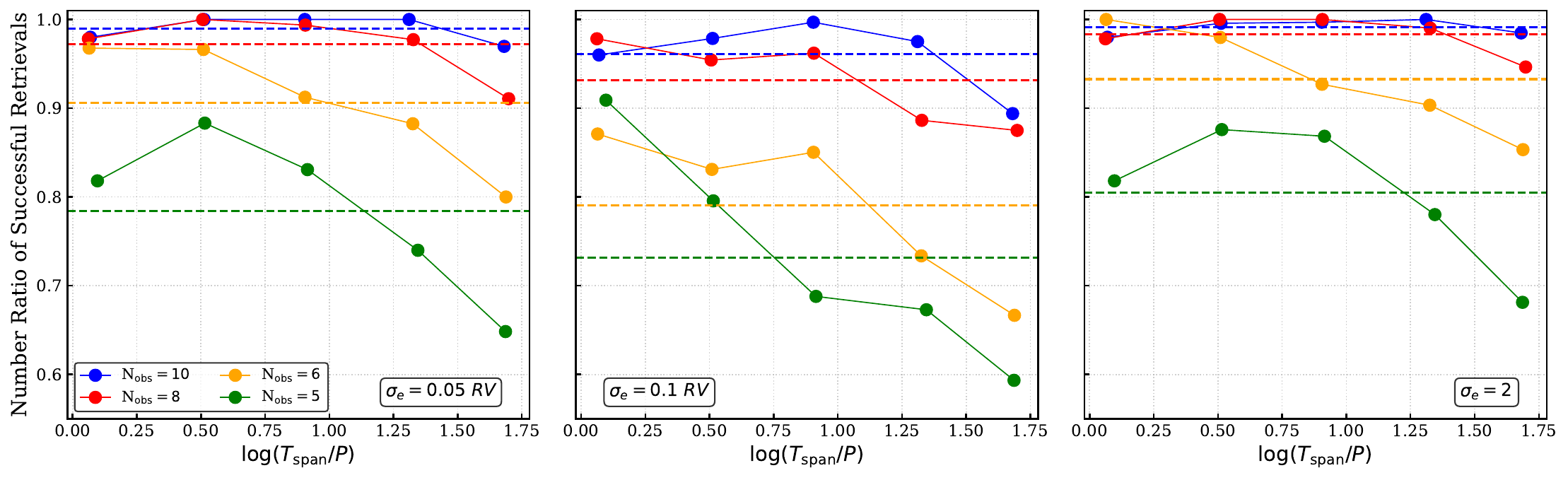}
    \includegraphics[width=0.9\linewidth]{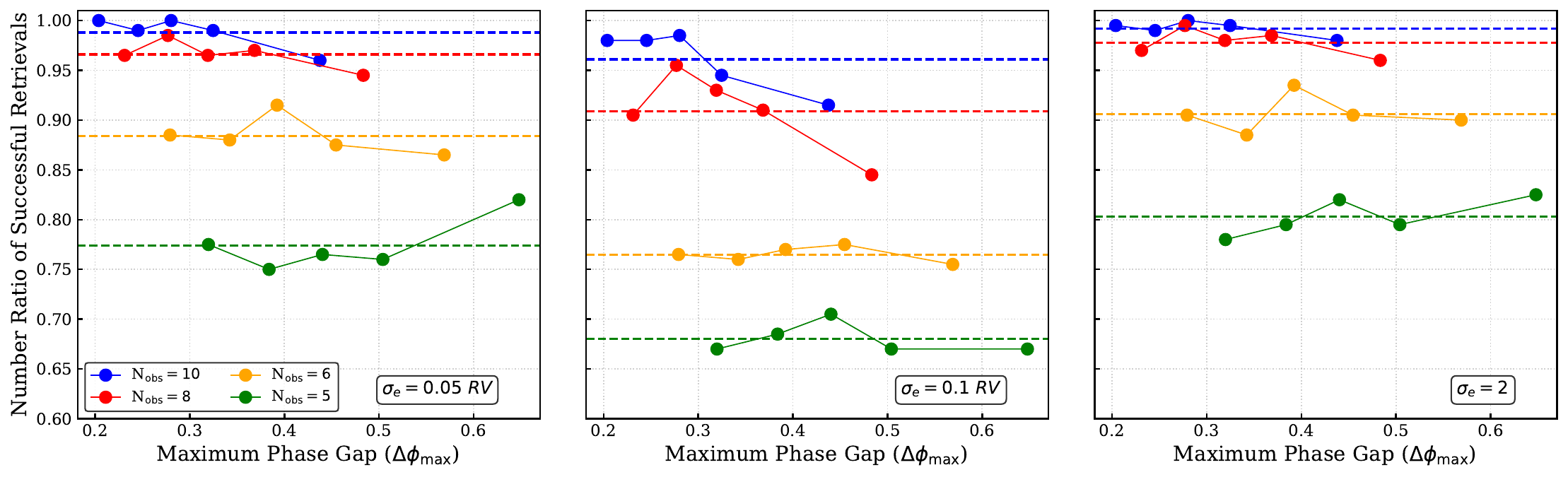}
    
    \caption{Period recovery success rates (within a 10\% accuracy threshold) for 1,000 synthetic binary systems under varying heteroscedastic noise levels (scaled to $\sigma/RV = 0.05$ and $0.1$, and fixed to $\sigma=2$, columns from left to right). Rows display performance as a function of $P$, $e$, $\log(T_{\rm span}/P)$, and $\Delta \phi_{\max}$ from top to bottom, where markers indicate the mean values of the given bin.   Colors correspond to different $N_{\rm obs}$ values, and horizontal dashed lines denote the overall mean success rate for each $N_{\rm obs}$ group at that specific noise level.}
    \label{fig:a1}
\end{figure}

Presenting all combinations of noise prescriptions and observational regimes would require a prohibitively large number of figures. We therefore illustrate representative cases using $\sigma/|RV|=0.05$, $\sigma/|RV|=0.1$, and $\sigma=2$, which approximately correspond to representative observational uncertainties, a high-noise regime, and a fixed-sigma noise scenario, respectively. The Figure~\ref{fig:a1} show the recovery fractions as functions of $P$, $e$, $T_{\rm span}/P$, and $\Delta\phi_{\max}$ for $N_{\rm obs}=10$, $8$, $6$, and $5$, together with the corresponding global recovery fractions.

Increasing the noise level gradually decreases the recovery fractions of both methods, particularly for sparse datasets. Nevertheless, the overall parameter-dependent trends remain qualitatively unchanged. Even for the low-noise case of $\sigma/|RV|=0.01$ and the extreme stress-test case of $\sigma/|RV|=0.2$, the distributions as functions of $P$, $e$, $T_{\rm span}/P$, and $\Delta\phi_{\max}$ preserve the same general morphology as in the noise-free simulations, although the absolute recovery fractions decrease with increasing uncertainty. These results indicate that the conclusions of Section~\ref{sec:test_results} remain robust against realistic measurement errors and even under considerably degraded observational conditions.

\bibliography{references}{}
\bibliographystyle{aasjournalv7}


\end{document}